\documentclass[journal]{IEEEtran}
\usepackage[dvips]{color}
\usepackage{amsmath}
\usepackage{amssymb}
\usepackage{mathrsfs}
\usepackage{graphicx}
\usepackage{upgreek}
\usepackage{bm}
\usepackage{algorithm}
\usepackage{algorithmic}
\usepackage{multirow,makecell}
\usepackage{subfigure}
\usepackage{xcolor}
\usepackage{CJKutf8}
\usepackage{cases}

\usepackage{cite}

\usepackage{xpatch}

\begin{document}

	\title{Movable Antenna-Enhanced Wireless Powered Mobile Edge Computing Systems}

			


\author{\IEEEauthorblockN{Pengcheng Chen, \emph{Student Member, IEEE}, Yuxuan Yang, \emph{Student Member, IEEE}, Bin Lyu, \emph{Member, IEEE}, \\ Zhen Yang, \emph{Senior Member, IEEE}, and Abbas Jamalipour, \emph{Fellow, IEEE}}

\IEEEcompsocitemizethanks{\IEEEcompsocthanksitem P. Chen, B. Lyu, and Z. Yang are with the Key Laboratory of Ministry of Education in Broadband Wireless Communication and Sensor Network Technology, Nanjing University of Posts and Telecommunications, Nanjing 210003, China (email: 2020010104@njupt.edu.cn, blyu@njupt.edu.cn, yangz@njupt.edu.cn).\protect

Y. Yang and A. Jamalipour are with the School of Electrical and Information Engineering, University of Sydney, Sydney, NSW 2006, Australia (email:yyan5256@uni.sydney.edu.au, a.jamalipour@ieee.org).   
}}

	\maketitle
	\begin{abstract}
		In this paper, we propose a movable antenna (MA) enhanced scheme for wireless powered mobile edge computing (WP-MEC) system, where the hybrid access point (HAP) equipped with multiple MAs first emits wireless energy to charge wireless devices (WDs), and then receives the offloaded tasks from the WDs for edge computing. The MAs deployed at the HAP enhance the spatial degrees of freedom (DoFs) by flexibly adjusting the positions of MAs within an available region, thereby improving the efficiency of both downlink wireless energy transfer (WPT) and uplink task offloading. To balance the performance enhancement against the implementation intricacy, we further propose three types of MA positioning configurations, i.e., dynamic MA positioning, semi-dynamic MA positioning, and static MA positioning. In addition, the non-linear power conversion of energy harvesting (EH) circuits at the WDs and the finite computing capability at the edge server are taken into account. Our objective is to maximize the sum computational rate (SCR) by jointly optimizing the time allocation, positions of MAs, energy beamforming matrix, receive combing vectors, and offloading strategies of WDs. To solve the non-convex problems, efficient alternating optimization (AO) frameworks are proposed. Moreover, we propose a hybrid algorithm of particle swarm optimization with variable local search (PSO-VLS) to solve the sub-problem of MA positioning. Numerical results validate the superiority of exploiting MAs over the fixed-position antennas (FPAs) for enhancing the SCR performance of WP-MEC systems.
	\end{abstract}
	\begin{IEEEkeywords}
		Movable antennas (MAs), wireless powered mobile edge computing (WP-MEC), MA positioning configurations, particle swarm optimization with variable local search (PSO-VLS).
	\end{IEEEkeywords}
	\section{Introduction}
  	With the rapid development of Internet of Things (IoT), the scale of data produced by heterogeneous wireless devices (WDs) has grown exponentially in recent years \cite{kong2022edge}. However, constrained by the limitations of physical dimensions, the batteries of WDs are normally capacity-constrained, resulting in insufferable replacement cost. In addition, for cost saving, the WDs are typically equipped with processors with poor computing capability. As such, these WDs can not sustainably power the emerging resource-intensive applications with strict delay demands \cite{XuSai2023Intelligent}. Cloud computing, allowing task offloading from WDs to the powerful cloud servers, can significantly amplify the computing capability of WDs. However, these servers are always deployed in internet center and far away from the WDs, which causes severe energy loss and unbearable response delay. 
  		
	Mobile edge computing (MEC) has emerged as an effective solution for enhancing the capabilities of WDs by placing servers in close proximity to the WDs \cite{SiriwardhanaASurvey}. Compared to the cloud computing, MEC facilitates more efficient access for WDs to the computation resources of the edge servers over the centralized cloud in the distance \cite{MaoYuyi2017ASurvey}. Wireless power transfer (WPT) technique has gained increasing attention for its ability of handling the critical issue of energy shortage at the WDs \cite{ZhangHaiyang2022NearField}. Specifically, the WDs featuring energy harvesting (EH) circuits can harvest the energy of radio frequency (RF) signals from a controllable energy station and then store it in rechargeable batteries \cite{ZhangZhen2019Wireless}. In this way, the batteries of the WDs keep on recharging, which makes the self-sustainability of WDs realistic. The combination of WPT and MEC, i.e., wireless powered MEC (WP-MEC), has pioneered an appealing paradigm for ubiquitous computing in a self-sustainable manner \cite{wang2023wireless}. In WP-MEC systems, WDs leverage the harvested energy for local computing and/or task offloading. The proliferation of resource-intensive applications has necessitated the utilization of multiple antennas at the energy station to achieve efficient WPT systems by virtue of favorable energy beams. However, the spatial degrees of freedom (DoFs) are not fully exploited since the positions of the conventional antennas remain stationary.	
	
	Recently, an emerging antenna paradigm called movable antenna (MA) has originated from the above dilemma, which is able to fully exploit such DoFs in the continuous spatial region \cite{zhu2023movable}. Specifically, in MA-enhanced systems, each antenna is linked with an RF chain through a flexible cable, whose length can be flexibly adjusted in a certain range. In addition, each antenna is equipped with a driver component, e.g., stepper motor or servo, and thus it can move within an available spatial region \cite{ZhuLipeng2023Modeling, MaWenyan2023MIMO, cheng2023sumrate}. Compared to the fixed-position antennas (FPAs), the flexible mobility of MAs facilitates the antennas at favorable positions for better channel conditions so as to realize efficient WPT and communications. Generally, fewer antennas are acquired by the MA-enhanced systems to utilize such spatial DoFs \cite{sun2023sumrate}, which substantially reduces the computational complexity of signal processing in comparison with the FPA-enabled systems. 

	\subsection{Related Works}
	We review the existing works from next three perspectives, i.e., resource allocation in MEC systems, designs of WP-MEC systems, and MA-enhanced communications, respectively.
		
	\subsubsection{Resource allocation in MEC systems}
	The objective of MEC is to merge wireless communications and mobile computing seamlessly, yielding diverse designs of offloading strategies and network architectures. Considering the load imbalance at the edge sides caused by the differences in tasks and network capability, the authors in \cite{Teng2023Game} designed an efficient task offloading architecture to realize flexible collaborations between multiple edge servers. The authors in \cite{Bahreini2022Mechanisms} developed an efficient pricing mechanisms for resource allocation based on auction and linear programming. The authors in \cite{ZhouWenqi2023Profit} studied the impacts of task priority of heterogeneous WDs in the unmanned aerial vehicle (UAV) mounted MEC systems, where the offloading strategies and UAV trajectory are jointly optimized according to the time-varying priorities of tasks. The authors in \cite{Asghari2022Multiobjective} studied the placements of services in MEC systems to reduce the response latency. An edge platform was constructed in \cite{GuimFrancesc2022Autonomous}, which allowed flexible resources allocation for WDs with diverse requirements. 
	
	\subsubsection{Designs of WP-MEC systems}
	To facilitate ubiquitous computing sustainability, WPT-enabled MEC, has attracted increasing attention. The authors in \cite{BiSuzhi2018Computation} studied a weighted SCR maximization problem in the WP-MEC system, where the WDs follow the binary offloading mode. The authors in \cite{MaoSun2021Computation} studied the resource management for IRS-assisted WP-MEC system, where the energy consumption of the IRS and WDs was considered. To mitigate the doubly near-far effects, the authors in \cite{LiBaogang2021Wireless} proposed a cooperative scheme, in which the far-WD with less harvested energy can offload its tasks with the aid of the near-WD with better channel condition. The authors in \cite{Feng2021Hybrid} utilized an UAV to provide wireless energy and receive offloaded tasks in the vicinity of WDs. The author in \cite{WuMengru2023Energyefficient} focused on the security of task offloading and leveraged the power station to emit jamming signals to interfere with the eavesdroppers. The authors in \cite{YeYinghui2021Delay} and  \cite{YeYinghui2022Resource} studied backscatter-aided WP-MEC systems, in which the WDs can simultaneously obtain computing services from the edge server and harvest energy. The authors in \cite{ShiLiqin2021ComputationBits} also studied a backsctter-aided WP-MEC system and maximized the weighted SCR of all the WDs. The authors in \cite{Han2022Joint} studied the problem of max-min energy fairness among the multiple WDs. In addition, the authors in and\cite{ChenPengcheng2023Computational, ChenPengcheng2022MultiIRS, ChenPengcheng2023ComputationalRate} maximized the computational rate in the WP-MEC systems, where the time spent on edge computing is considered. 
	
	\subsubsection{MA-enhanced communications}
	MA has emerged as a promising approach to achieving better communication performance by exploiting new spatial DoFs via flexibly adjusting the positions of MAs. The authors in \cite{ZhuLipeng2023Modeling} developed a field-response based channel model under the far-field conditions and analyzed the maximum channel gain of the MA-aided system, in which both the receiver and transmitter are equipped with a single MA. For further capacity improvement, the authors in \cite{MaWenyan2023MIMO} studied an MA-aided point-to-point multiple-input multiple-output (MIMO) system, and jointly optimized the positions of MAs at both receiver and transmitter. The authors in \cite{cheng2023sumrate} and \cite{sun2023sumrate} studied the MA-enabled multiuser communication networks and maximized the sum-rate of all the WDs.	To reduce the pilot overhead, the authors in \cite{Xiao2023Channelestimation} and \cite{ma2023compressedSensing} proposed a compressed sensing based channel estimation framework, by which the complete channel state information (CSI) can be reconstructed via limited number of channel measurements. The authors in \cite{ChenXintai2023Joint} maximized the achievable rate by jointly optimizing the MA positions and the transmit covariance matrix based on statistical CSI. The authors in \cite{MaWenyan2024MultiBeam} maximized the minimum beamforming gain over the desired directions by joint optimizing the positions and  weight vector of the MA array. The effectiveness of the directional gain of the linear MA array was confirmed in \cite{ZhuLipeng2023MovableCL}, i.e., it enhances the gain in the target direction and suppresses interference in non-target directions. Moreover, the authors in \cite{xiao2023multiuser} maximized the minimum capability among all the WDs in an MA-aided multi-user system, and further validated the superiority of MAs in channel gain enhancement and interference suppression over the FPAs. The authors in \cite{zhu2023movablevia, hu2023movable, qin2023antenna} demonstrated that MAs have significant power saving properties. Moreover, the benefits of MAs in physical security are exploited and conformed by the authors in \cite{hu2023secure}. 
	
	\subsection{Motivations and Contributions}	
	In the previous studies (i.e., \cite{BiSuzhi2018Computation, MaoSun2021Computation, ChenPengcheng2023ComputationalRate, YeYinghui2022Resource, ShiLiqin2021ComputationBits, ChenPengcheng2023Computational, ChenPengcheng2022MultiIRS, Han2022Joint, LiBaogang2021Wireless, YeYinghui2021Delay, Feng2021Hybrid, WuMengru2023Energyefficient}), the transmitters and receivers in WP-MEC systems were equipped with the conventional FPAs, which cannot fully exploit spatial diversity gains. MA technology, flexibly deploying the antennas at the positions with better channel conditions, has been regarded as a novel approach to acquiring the spatial DoFs. Therefore, it is necessary to exploit the advantages of MA technology in the WP-MEC systems for further performance improvement.
	
    Recently, several works have validated the advantages of MA in capacity improvement \cite{ZhuLipeng2023Modeling, MaWenyan2023MIMO, cheng2023sumrate, sun2023sumrate}, interference suppression \cite{ZhuLipeng2023MovableCL, xiao2023multiuser}, power conservation \cite{zhu2023movablevia, hu2023movable, qin2023antenna}, and physical security \cite{hu2023secure} in communications. However, the methods proposed in \cite{ZhuLipeng2023Modeling, MaWenyan2023MIMO, cheng2023sumrate, sun2023sumrate} and \cite{ZhuLipeng2023MovableCL, xiao2023multiuser,zhu2023movablevia, hu2023movable, qin2023antenna,hu2023secure} are not suitable for the WP-MEC systems. It is due to the fact that the MA technology has not been used for the WPT improvement to the best of our knowledge. Hence, it is necessary to work out a new method with high accuracy to optimizing the positions of MAs in the WP-MEC systems.
	
	In addition, the authors in \cite{BiSuzhi2018Computation, MaoSun2021Computation, LiBaogang2021Wireless, Feng2021Hybrid, WuMengru2023Energyefficient, YeYinghui2021Delay, YeYinghui2022Resource, Han2022Joint, ChenPengcheng2023Computational, ChenPengcheng2022MultiIRS, ShiLiqin2021ComputationBits} assumed that the edge server executes the offloaded tasks after the task offloading stage. As a result, the edge server keeps idle for a long time during each time block. Moreover, the previous works (i.e., \cite{BiSuzhi2018Computation, MaoSun2021Computation, LiBaogang2021Wireless, Feng2021Hybrid, WuMengru2023Energyefficient, YeYinghui2021Delay, ShiLiqin2021ComputationBits} and \cite{Han2022Joint}) ideally assumed that the computing capability of edge server is infinite and ignore the time spent on executing the offloaded tasks.

	Based on above observations, we propose an MA-enhanced WP-MEC system, where the HAP is equipped with MA array instead of the conventional FPAs. The MA-enabled HAP emits wireless energy to charge the WDs firstly and then receives the offloaded tasks from the WDs. The WPT efficiency can be improved and the communication links for task offloading can be strengthened by properly adjusting both the beamforming designs and the positions of MA array at the HAP.Due to the non-linear conversion efficiency of the built-in rectifier, the EH circuit typically causes a non-linear power conversion \cite{Jiapin2012Animproved}. To capture the non-linear power conversion of EH circuits at the WDs, a practical EH model is applied. In addition, the finite computing capability of edge server is considered. According to the characteristics of the partial offloading mode (i.e., arbitrary segmentation and parallel computation \cite{MaoYuyi2017ASurvey}), we consider the immediate response at the edge server, i.e., the edge server immediately executes the offloaded tasks as soon as they are received. In this way, the finite edge computing capability can be fully exploited. Our objective is to maximize the sum computational rate (SCR) of all WDs by jointly optimizing the time allocation, the positions of MAs, the energy beamforming matrix and receive combing vectors at the HAP, as well as the offloading strategies of the WDs. The main contributions are summarized as follows.
	\begin{itemize}
		\item To the best of our knowledge, this is the first work that exploits MAs in the WP-MEC systems. To balance the enhancement of performance against the augmentation in implementation intricacy due to the MA positioning optimization, three types of MA positioning configurations, i.e., dynamic MA positioning, semi-dynamic MA positioning, and static MA positioning, are proposed based on how the MAs are allowed to adjust their positions across time. Specifically, for the dynamic MA positioning scheme, the positions of MAs can be adjusted flexibly not only for WPT but also for each WD's task offloading, thereby improving the WPT efficiency and strengthening the communication links significantly. To achieve a better trade-off between the performance enhancement and the implementation intricacy, the semi-dynamic MA positioning scheme allows all WDs to share the same positions of MAs during the entire task offloading stage. In addition, the implementation intricacy can be further reduced by the static MA positioning scheme, in which the positions of MAs remain stationary throughout the entire time block.
		
		\item For the case with dynamic MA positioning, we propose an efficient alternating optimization (AO) framework to solve the formulated problem. Specifically, the feasibility-checking sub-problem of energy beamforming matrix and MA positioning optimization during the WPT stage is firstly converted into an explicit goal-based problem. Then, the successive convex approximation (SCA) is applied to tackle the non-convexity caused by the non-linear EH model. In addition, according to the maximum-ratio combining (MRC) criterion, the receive combing vectors are calculated in the closed-form with respect to the positions of MAs. The hybrid algorithm of particle swarm optimization with variable local search (PSO-VLS) is proposed to optimize the positions of MAs. Compared to the standard PSO algorithm in \cite{xiao2023multiuser}, the PSO-VLS algorithm can achieve a better SCR performance.
		
		\item For the case with semi-dynamic MA positioning and the case with static MA positioning, we extend the AO framework with the PSO-VLS algorithm to solve the formulated problems. Specifically, the MA positioning optimizations for both cases are solved by the PSO-VLS algorithm by properly redesigning the fitness functions.
		
		\item Numerical results validate the high efficiency of the PSO-VLS algorithm and the AO frameworks. Compared to the standard PSO algorithm, the proposed PSO-VLS algorithm provides solutions for MA positioning with higher accuracy, especially for the scenario with sufficient channel paths. Moreover, extensive results demonstrate the superiority of the MA-enhanced scheme over the FPA-enabled scheme in WP-MEC systems. 
	\end{itemize}
		
	The rest of this paper is organized as follows.
	Section \ref{section_system_model_and_problem_formulation} introduces the system model and problem formulations for the MA-enhanced WP-MEC system with three types of MA positioning configurations. Section \ref{section_solution_for_dynamic_MA_positioning} proposes an efficient AO framework to solve the formulated problem for the case with dynamic MA positioning. In Section \ref{section_solution_for_semi-dynamic_MA_positioning and Static MA Positioning}, the AO framework is extended to solve the formulated problems for the cases with semi-dynamic MA positioning and static MA positioning, respectively. Section \ref{section_numerical_results} provides the numerical results. Section \ref{section_conclusion} concludes this paper.

	\section{System Model and Problem Formulation} 
	\label{section_system_model_and_problem_formulation}
	\begin{figure}[t]
		\centering\includegraphics[width=8.5cm]{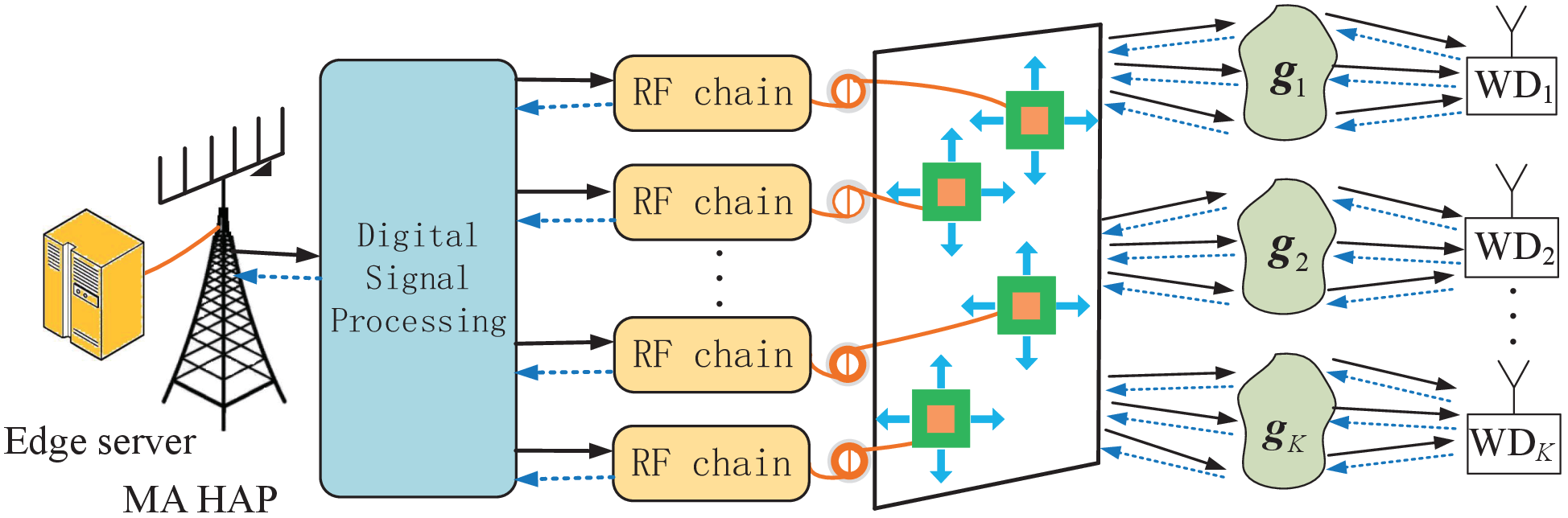}
		\caption{The MA-enhanced WP-MEC system.}
		\label{fig_system_model}
	\end{figure}
	\begin{figure}[t]
		\centering\includegraphics[width=8.5cm]{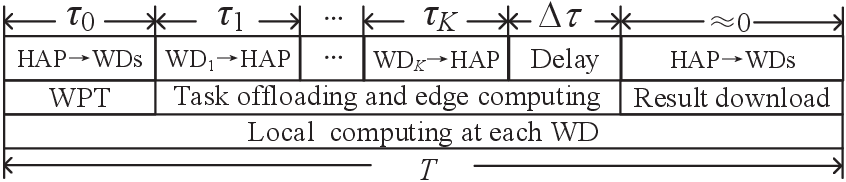}
		\caption{The time scheduling.}
		\label{fig_frame_structure}
	\end{figure}
	As depicted in Fig.~\ref{fig_system_model}, we consider an MA-enhanced WP-MEC system, which comprises an HAP with $M$ MAs, an edge server with finite computing capability, and $K$ single-FPA WDs. Each of the MAs at HAP is connected to an RF chain via a flexible cable and is driven by a stepper motor \cite{ZhuLipeng2023Modeling}. Thus, the positions of MAs can be mechanically adjusted within a given region at the HAP, i.e.,  $\mathcal{C}$. The HAP firstly emits wireless energy to charge the WDs and then receives the offloaded tasks from the WDs. The WDs are assumed to operate in the partial offloading mode, i.e., offloading tasks while processing computational tasks locally. To guarantee completely execution of the offloaded tasks at the edge server, the processing delay caused by the finite computing capability of edge server needs to be taken into account. 
	
	Fig.~\ref{fig_frame_structure} illustrates the scheduling of the considered time block with duration of $T$ in seconds, which consists of the WPT stage, the task offloading and edge computing stage, as well as the result downloading stage. The WPT stage with duration of $\tau_{0}$ is used for charging the WDs. Then, during the task offloading and edge computing stage, the WDs may opt for employing part of the harvested energy to offload some tasks to the edge server in rotation. Specifically, the slot with duration of $\tau_{k}$ is exclusively used for WD $k$ to offload its tasks, where $k\in\mathcal{K}\triangleq\{1,\cdots,K\}$. To fully exploit the finite edge computing capability, we consider the immediate response at the MEC server, i.e., the MEC server immediately processes any offloaded tasks from WDs once they are received, instead of implementing the edge computing after completing the offloading of all tasks \cite{YeYinghui2022Resource}. In addition, a dedicated slot with duration of $\Delta\tau$ is required to guarantee the complete execution of residual tasks. Similar to \cite{BaiTong2021Resource}, the duration for result downloading can be negligible due to the minuscule volume of results typically involved. Advanced power splitting techniques empower the on-chip processors of WDs during the WPT stage by utilizing portion of the harvested energy, facilitating continuous local computing throughout the whole time block \cite{Mishra2018Transmit}.
	
	In this paper, three types of MA positioning configurations, i.e., dynamic MA positioning, semi-dynamic MA positioning, and static MA positioning, are considered. Without loss of generality, the case with dynamic MA positioning is modeled in details. As mentioned above, the positions of MAs can be flexibly adjusted during each time slot in this case.
	
	\subsection{Channel Model}
	Similar to \cite{ZhuLipeng2023Modeling, MaWenyan2023MIMO, ZhuLipeng2023MovableCL, xiao2023multiuser}, The field-response channel model is employed, i.e, the channel response is the superposition of coefficients of multiple channel paths between the HAP and WDs. In addition, for this MA-enhanced WP-MEC system, the far-field condition holds due to the smaller moving area for MAs compared with signal transmission range between the WDs and the HAP. Thus, the plane-wave model can be utilized to form the field response channel \cite{ZhuLipeng2023Modeling}. Specifically, the angle of departure (AoD), the angle of arrival (AoA), and the amplitude of complex coefficient for each channel path remains constant despite various positions of MAs, while only the phases of the channel paths various with respect to the positions of MAs.
	
	Let Cartesian coordinates $\bm{r}_{i,m}=[x_{i,m},y_{i,m}]$ denote the position of the $m$-th MA during $\tau_{i}$, where $i\in\mathcal{I}\triangleq\{0,1,\cdots,K\}$ and $m\in\mathcal{M}\triangleq\{1,\cdots,M\}$. Let $L_{k}$ denote the number of channel paths between the HAP and WD $k$. The elevation and azimuth AoDs of the $l$-th channel path at the HAP for WD $k$ are respectively denoted by $\theta_{k,l}\in [0,\pi]$ and $\phi_{k,l}\in [0,\pi]$. Then, for this channel path, the deference of signal propagation  between the position $\bm{r}_{i,m}$ and the reference point $\bm{r}_{0}=[0,0]$ is 
	\begin{equation}
		\rho_{k,l}(\bm{r}_{i,m}) = x_{i,m}\mathrm{sin}(\theta_{k,l})\mathrm{cos}(\phi_{k,l}) + y_{i,m}\mathrm{cos}(\theta_{k,l}).
	\end{equation}
	Accordingly, during $\tau_{i}$, the field response vector (FRV) of between the $m$-th MA and WD $k$ is obtained as 
	\begin{equation*}
		\bm{f}_k(\bm{r}_{i,m}) = \big[e^{j\tfrac{2\pi}{\lambda}\rho_{k,1}(\bm{r}_{i,m})},\cdots,e^{j\tfrac{2\pi}{\lambda}\rho_{k,L_{k}}(\bm{r}_{i,m})}\big]^{T}\in\mathbb{C}^{L_k\times 1},
	\end{equation*}
	where $\lambda$ is the carrier wavelength.
	By stacking $\bm{f}_k(\bm{r}_{i,m})$ of all MAs, the field response matrix (FRM) from WD $k$ and the $m$-th MA during $\tau_{i}$ is 
	\begin{equation}
		\mathbf{F}_{i,k}(\widetilde{\bm{r}}_{i}) \triangleq \left[\bm{f}_k(\bm{r}_{i,1}),\cdots, \bm{f}_k(\bm{r}_{i,M})\right]\in\mathbb{C}^{L_k\times M},
	\end{equation}
	where $\widetilde{\bm{r}}_{i}\triangleq[\bm{r}_{i,1},\cdots,\bm{r}_{i,M}]$ denote the antenna position vector (APV) of all MAs during $\tau_{i}$.
	As such, the channel response vector (CRV) from WD $k$ to the HAP is  
	\begin{equation}
		\bm{h}_k(\widetilde{\bm{r}}_{i}) = \mathbf{F}_{i,k}^{H}(\widetilde{\bm{r}}_{i})\bm{g}_{k}\in\mathbb{C}^{M\times 1},
	\end{equation}
	where $\bm{g}_{k}=[g_{k,1},\cdots,g_{k,l},\cdots,g_{k,L_k}]^T\in\mathbb{C}^{L_k\times 1}$ is the path response vector (PRV). Herein, $g_{k,l}$ is assumed to be block-fading and perfectly estimated by the methods proposed in \cite{Xiao2023Channelestimation} and \cite{ma2023compressedSensing}.
	
	The channel reciprocity between the downlink channels and the counterpart uplink channels is assumed and thus the CRV from the HAP and WD $k$ during $\tau_{i}$ is $\bm{h}_k^T(\widetilde{\bm{r}}_{i})$ \cite{LiZhendong2022Robust}.

	\subsection{Wireless Power Transfer}
	Let $\bm{x}_0=\sum\limits_{j=1}^{M_t}\bm{w}_{0,j}x_{0,j}$ denote the combined effect of multiple independent energy beams emitted by the HAP, where $M_t\leq M$ is the number of energy beams, and $x_{0,j}$ is the carried signal satisfying $\mathbb{E}][\vert x_{0,j}\vert^2]=1$. Accordingly, during $\tau_0$, the signal received by WD $k$ is
	\begin{equation}
		y_{0,k} = \bm{h}_k^T(\widetilde{\bm{r}}_{0})\bm{x}_0 + n_k,
	\end{equation}
	where $n_k\sim\mathcal{CN}(0,\sigma_0^2)$ denotes the additive white Gaussian noise (AWGN). For simplicity, let $\mathbf{Q} \triangleq \mathbb{E}\left[\Vert\bm{x}_0\Vert^2\right] = \sum\limits_{j=1}^{M_t}\bm{w}_{0,j}\bm{w}_{0,j}^H\succeq0$ denote the energy beamforming matrix. Mathematically, $\bm{w}_{0,j}$ can be recovered by the eigenvalue decomposition (EVD) of $\mathbf{Q}$.

   To capture the non-linear power conversion of the EH circuits, a practical non-linear EH model is applied \cite{HuaMeng2022Throughput}. Accordingly, the amount of power harvested by WD $k$ is  
	\begin{equation*}
		\Xi_k=\dfrac{X_k}{1+\textrm{exp}\left(-a_k\Big(\bm{h}_k^T(\widetilde{\bm{r}}_0)\mathbf{Q}\big(\bm{h}_k^T(\widetilde{\bm{r}}_0)\big)^H-b_k\Big)\right)}-Y_k.
	\end{equation*} 
	Herein, $a_k$, $b_k$,  $M_k$,  $X_k={M_k\big(1+\textrm{exp}(a_kb_k)\big)}/{\textrm{exp}(a_kb_k)}$, and $Y_k={M_k}/{\textrm{exp}(a_kb_k)}$ are constants determined by the configurations of the EH circuit.

	\subsection{Computing Model}
	\subsubsection{Local Computing}
	As mentioned above, the local computing rate of WD $k$ in bits/second (bps) is
	\begin{equation} \label{Rate_local}
		R_{L,k}=\dfrac{f_kT}{\varphi_kT}=\dfrac{f_k}{\varphi_k},
	\end{equation}
	where $f_k$ (Hz) denote the frequency of WD $k$, and $\varphi_k$ (cycles/bit) denotes the complexity of the tasks at WD $k$. 
	The energy consumption for local computing is 
	\begin{equation}
		E_{L,k}=\kappa f_k^3T,
	\end{equation}
	where $\kappa$ ($\mathrm{Watt/Hz^3}$) is the capacitance coefficient of the on-chip processor at WD $k$ \cite{wang2016mobile}.
	
	\subsubsection{Task Offloading}
	The WDs may opt for performing task offloading in rotation by employing part of the harvested energy. Let $x_k$ denote the transmit signal of WD $k$ satisfying $\mathbb{E}\left[\vert x_k\vert^2\right]=1$.  The estimated signal at the HAP is   
	\begin{equation}
		y_k=\bm{w}_k^H\bm{h}_k(\widetilde{\bm{r}}_k)\sqrt{p_k}s_k+\bm{w}_k^H\bm{n}_0,
	\end{equation} 
	where $\bm{w}_k\in\mathbb{C}^{M\times1}$ is a receive combing vector to combine the received signal at the HAP for WD $k$, $p_k$ is the transmit power of WD $k$, and $\bm{n}_0\sim\mathcal{CN}(\bm{0},\sigma_0^2\bm{I}_M)$ is the AWGN. 
	The offloading rate of WD $k$ is
	\begin{equation} \label{Rate_offload}
		R_{O,k} = \dfrac{B\tau_k}{T}\log_2\Big(1+\dfrac{|\bm{w}_k^H\bm{h}_k(\widetilde{\bm{r}}_k)|^2p_k}{\Vert\bm{w}_{k}^H\Vert^2\sigma_0^2}\Big),
	\end{equation}
	where $B$ (Hz) is the frequency bandwidth. 
	
	\subsubsection{Edge Computing}
	Similar to \cite{ChenPengcheng2023ComputationalRate}, to guarantee all offloaded tasks are impeccably executed by edge servers, we have
	\begin{equation} 
		\sum_{i=k}^{K}\varphi_iR_{O,i}T\leq f_E \cdot \Big(\Delta\tau+\sum_{i=k}^{K}\tau_{i}\Big),
	\end{equation}
	where $f_E$ (Hz) denotes the edge computing frequency. 
	
	\subsection{Problem Formulation}
	Denote by sets $\bm{\tau}=\{\Delta\tau,\tau_i,i\in\mathcal{I}\}$,  $\bm{p}=\{p_k,k\in\mathcal{K}\}$, $\bm{f}=\{f_k,k\in\mathcal{K}\}$,$\mathbf{w}=\{\bm{w}_k,k\in\mathcal{K}\}$, and  $\widetilde{\mathbf{r}}=\{\widetilde{\bm{r}}_i,i\in\mathcal{I}\}$. Our objective is to maximize the SCR of all WDs in the MA-enhanced WP-MEC system by jointly optimizing the time allocation, positions of MAs, energy beamforming matrix, receive combing vectors, and offloading strategies of WDs. 
	\subsubsection{Dynamic MA positioning}
	In this case, the optimization problem is formulated as
	\begin{subequations} 
		\begin{align}
			(\textrm{PA}):\ &\max_{\bm{\tau},\bm{\bm{p}},\bm{f},\mathbf{w},\widetilde{\mathbf{r}},\mathbf{Q}\succeq0}\sum_{k\in\mathcal{K}}(R_{L,k}+R_{O,k}) \label{Obj_for_dynamic_MA_positioning} \\ 
			s.t.\ 
			& \kappa f_k^3T + p_k\tau_k \leq \tau_0\Xi_k, \ k\in\mathcal{K}, \label{Con_energy_causality} \\
			& \sum_{i\in\mathcal{I}}\tau_{i} + \Delta\tau \leq T,\ 0\leq\tau_i,\Delta\tau\leq T, \label{Con_time_allocation} \\ 
			& \sum_{i=k}^{K}\varphi_iR_{O,i}T\leq f_E \cdot \Big(\Delta\tau+\sum_{i=k}^{K}\tau_{i}\Big), \ k\in\mathcal{K} \label{Con_edge_computing},\\
			& \mathrm{Tr}(\mathbf{Q})\leq P_\textrm{max}, \label{Con_energy_beamforming_matrix}\\ 
			& \bm{r}_{i,m}\in\mathcal{C},\ m\in\mathcal{M}, i\in\mathcal{I},  \label{Con_MA_range} \\
			& \Vert\bm{r}_{i,m}-\bm{r}_{i,n}\Vert_2 \geq D,\ 1\leq m\neq n\leq M, i\in\mathcal{I}, \label{Con_MA_distance}
		\end{align}
	\end{subequations}
	where $P_\textrm{max}$ is the maximum transmit power of the HAP.
	(\ref{Con_energy_causality}) is the energy-causality constraint, which indicates that the energy consumed by each WD cannot exceed the harvested energy.  (\ref{Con_edge_computing}) is caused by the finite edge computing capability. (\ref{Con_MA_distance}) ensures that the inter-MA distance should be not less than $D$ in actual practice. 
	
	\subsubsection{Semi-dynamic MA positioning}
	In this case, two APVs are allocated for WPT and task offloading, respectively. Let $\widetilde{\bm{r}}_u$ denote the APV during entire task offloading stage,  i.e., $\widetilde{\bm{r}}_u=\widetilde{\bm{r}}_k$, $\forall k\in\mathcal{K}$. Then, the problem is formulated as 
	\begin{subequations} \label{problem_semi_dysnamic_MA_positioning}
		\begin{align}
			(\textrm{PB}):\ &\max_{\bm{\tau},\bm{\bm{p}},\bm{f},\mathbf{w},\widetilde{\bm{r}}_0,\widetilde{\bm{r}}_u,\mathbf{Q}\succeq0}\sum_{k\in\mathcal{K}}(R_{L,k}+R_{O,k}) \\ 
			&s.t.\ 
			 (\textrm{\ref{Con_energy_causality}}),(\textrm{\ref{Con_time_allocation}}),(\textrm{\ref{Con_edge_computing}}),(\textrm{\ref{Con_energy_beamforming_matrix}}), \notag\\ 
			&\qquad \bm{r}_{i,m}\in\mathcal{C},\ m\in\mathcal{M},i\in\{0,u\}, \label{Con_MA_range_semi_dynamic}  \\
			\Vert\bm{r}_{i,m}&-\bm{r}_{i,n}\Vert_2 \geq D,\ 1\leq m\neq n\leq M, i\in\{0,u\}. \label{Con_MA_distance_semi_dynamic} 
		\end{align}
	\end{subequations}
	
	\subsubsection{Static MA positioning}
	In this case, the same APV, denoted by $\widetilde{\bm{r}}_s$, is adopted throughout the entire time block, i.e., $\widetilde{\bm{r}}_s=\widetilde{\bm{r}}_i$, $\forall i\in\mathcal{I}$. Then, the problem is formulated as 
	\begin{subequations} 
		\begin{align}
			(\textrm{PC}):\ &\max_{\bm{\tau},\bm{\bm{p}},\bm{f},\mathbf{w},\widetilde{\bm{r}}_s,\mathbf{Q}\succeq0}\sum_{k\in\mathcal{K}}(R_{L,k}+R_{O,k}) \\ 
			s.t.\ & (\textrm{\ref{Con_energy_causality}}),
			(\textrm{\ref{Con_time_allocation}}), (\textrm{\ref{Con_edge_computing}}),  (\textrm{\ref{Con_energy_beamforming_matrix}}), \notag  \\ 
			& \bm{r}_{s,m}\in\mathcal{C},\ m\in\mathcal{M}, \label{Con_MA_range_static}  \\
			& \Vert\bm{r}_{s,m}-\bm{r}_{s,n}\Vert_2 \geq D,\ 1\leq m\neq n\leq M.\label{Con_MA_distance_static} 
		\end{align}
	\end{subequations}

	Obviously, the above three problems are all highly non-convex due to the non-convexity of objective functions with APVs, as well as the multiplicative terms in $\textrm{(\ref{Con_energy_causality})}$ and $\textrm{(\ref{Con_edge_computing})}$. The globally optimal solutions can not be obtained by the existing convex optimization tools. 
	
	\section{Solution for Dynamic MA Positioning}
	\label{section_solution_for_dynamic_MA_positioning}
	In this section, we propose an efficient AO framework to obtain a sub-optimal solution for (PA). Specifically, the variables are first partitioned as $\{\bm{\tau},\bm{p},\bm{f}\}$, $\{\widetilde{\bm{r}}_0,\mathbf{Q}\}$, and $\{\mathbf{w},\widetilde{\bm{r}}_k,k\in\mathcal{K}\}$. Then, we optimize each block of variables alternately, until convergence is achieved.
	
	\subsection{Optimization of $\{\bm{\tau}, \bm{p}, \bm{f}\}$}
	Given $\{\widetilde{\bm{r}}_0,\mathbf{Q}\}$ and $\{\mathbf{w}, \widetilde{\bm{r}}_k, k\in\mathcal{K}\}$, (PA) is reduced to
	\begin{subequations} 
		\begin{align}
			(\textrm{PA1}):&\ \max_{\bm{\tau},\bm{p},\bm{f}} \sum_{k\in\mathcal{K}}(R_{L,k}+R_{O,k}) \notag \\
			&s.t.\ (\textrm{\ref{Con_energy_causality}}),			 (\textrm{\ref{Con_time_allocation}}),(\textrm{\ref{Con_edge_computing}}). \notag 
		\end{align}
	\end{subequations}

	According to \cite{ChenPengcheng2023Computational}, the maximum SCR can be obtained if each WD exhausts its harvested energy, i.e., 
	\begin{equation} \label{equation_energy_exhaust}
		\kappa f_k^3T + p_k\tau_{k} = \tau_0\Xi_k.
	\end{equation} 
	Then, $R_{L,k}$ in (\ref{Rate_local}) can be rewritten as 
	\begin{equation} \label{R_local_temp}
		R_{L,k}(\tau_0,p_k,\tau_{k})=\dfrac{1}{\varphi_k}\Big(\dfrac{\tau_0\Xi_k-p_k\tau_{k}}{\kappa T}\Big)^{\tfrac{1}{3}},
	\end{equation}
	which is non-convex with respect of $\tau_k$ and $p_k$. Then, the auxiliary variable $e_k=p_k\tau_k$ is introduced. Accordingly, $R_{L,k}(\tau_0,p_k,\tau_{k})$ in (\ref{R_local_temp}) and $R_{O,k}$ in (\ref{Rate_offload}) can be rewritten as 
	\begin{equation*}
		R_{L,k}(\tau_0,e_k)=\dfrac{1}{\varphi_k}\Big(\dfrac{\tau_0\Xi_k-e_k}{\kappa T}\Big)^{\tfrac{1}{3}},
	\end{equation*}
	and
	\begin{equation*}
		R_{O,k}(\tau_k,e_k) = \dfrac{B\tau_k}{T}\log_2\Big(1+\dfrac{e_k|\bm{w}_k^H\bm{h}_k(\widetilde{\bm{r}}_k)|^2}{\tau_{k}\Vert\bm{w}_{k}^H\Vert^2\sigma_0^2}\Big),
	\end{equation*}
	respectively. Then, (PA1) can be converted to a convex problem as follows:
	\begin{equation*}
		\begin{aligned}
			(\textrm{PA1.1}):\ & \max_{\bm{\tau},\bm{e}}\  \sum_{k\in\mathcal{K}}\big(R_{L,k}(\tau_0,e_k)+R_{O,k}(\tau_k,e_k)\big) \\
			&s.t.\  (\textrm{\ref{Con_time_allocation}}), (\textrm{\ref{Con_edge_computing}}), 0\leq e_k\leq \tau_{0}\Xi_k,\ k\in\mathcal{K},
		\end{aligned}
	\end{equation*}
	where $\bm{e}=\{e_k,k\in\mathcal{K}\}$.

	\subsection{Optimization of $\{\widetilde{\bm{r}}_0,\mathbf{Q}\}$} 
	\label{section_optimizing_MA_positioning_and_energy_beamforming_matrix_during_WPT_stage}
	Given $\{\bm{\tau},\bm{p},\bm{f}\}$ and $\{\mathbf{w},\widetilde{\bm{r}}_k,k\in\mathcal{K}\}$,  (PA) is reduced to a feasibility-checking problem as follows:
	\begin{subequations} 
		\begin{align}
			(\textrm{PA2}):\ &\textrm{Find}\ \{\widetilde{\bm{r}}_0,\mathbf{Q}\succeq 0\} \\ 
			s.t.\ 
			& (\textrm{\ref{Con_energy_causality}}),(\textrm{\ref{Con_energy_beamforming_matrix}}), \notag \\ 
			& \bm{r}_{0,m}\in\mathcal{C}, \ m\in\mathcal{M},  \label{Con_MA_range_0} \\
			& \Vert\bm{r}_{0,m}-\bm{r}_{0,n}\Vert_2 \geq D,\ 1\leq m\neq n\leq M. \label{Con_MA_distance_0}
		\end{align}
	\end{subequations}

	To solve (\textrm{PA2}) efficiently, it is better to convert (PA2) into an explicit goal-based problem. According to (\ref{equation_energy_exhaust}), auxiliary variable $\beta_k$ is introduced as energy partition factor. Specifically, part of the harvested energy, i.e., $\beta_k\tau_{0}\Xi_k$, is allocated for task offloading, and the residual of harvested energy, i.e., $(1-\beta_k)\tau_{0}\Xi_k$, is allocated for local computing at WD $k$. 
	Then, $R_{L,k}$ in (\ref{Rate_local}) and $R_{O,k}$ in (\ref{Rate_offload}) can be rewritten as 
	\begin{equation*} 
		R_{L,k}(\widetilde{\bm{r}}_0,\mathbf{Q}) = \dfrac{1}{\varphi_k}\Big(\dfrac{(1-\beta_k)\tau_{0}\Xi_k}{\kappa T}\Big)^{\tfrac{1}{3}}, 
	\end{equation*}
	and
	\begin{equation*} 
		R_{O,k}(\widetilde{\bm{r}}_0,\mathbf{Q}) = \dfrac{B\tau_k}{T}\log_2\Big(1+\dfrac{\beta_k\tau_{0}\Xi_k|\bm{w}_k^H\bm{h}_k(\widetilde{\bm{r}}_k)|^2}{\tau_{k}\Vert\bm{w}_{k}^H\Vert^2\sigma_0^2}\Big),
	\end{equation*}
	respectively. Then, (PA2) is converted into 
	\begin{subequations}
		\begin{align}
			(\textrm{PA2.1}):\ &\max_{\widetilde{\bm{r}}_0,\mathbf{Q}\succeq0}\sum_{k\in\mathcal{K}}\big(R_{L,k}(\widetilde{\bm{r}}_0,\mathbf{Q})+R_{O,k}(\widetilde{\bm{r}},\mathbf{Q})\big) \notag \\  
			&\qquad s.t.\ (\textrm{\ref{Con_edge_computing}}),  (\textrm{\ref{Con_energy_beamforming_matrix}}),(\textrm{\ref{Con_MA_range_0}}),(\textrm{\ref{Con_MA_distance_0}}), \notag 
		\end{align}
	\end{subequations}
	Obviously, (PA2.1) is non-convex since $\widetilde{\bm{r}}_0$ and $\mathbf{Q}$ are highly coupled. Then, (PA2.1) is further split  into two sub-problems with respect to $\widetilde{\bm{r}}_0$ and $\mathbf{Q}$, respectively.
	\subsubsection{Optimizing $\widetilde{\bm{r}}_0$} 
	\label{section_optimizating_r_0}
	Given $\mathbf{Q}$, (PA2.1) is reduced to
	\begin{subequations}
		\begin{align}
			(\textrm{PA2.2}):\ \max_{\widetilde{\bm{r}}_0}&\sum_{k\in\mathcal{K}}\big(R_{L,k}(\widetilde{\bm{r}}_0)+R_{O,k}(\widetilde{\bm{r}}_0)\big) \notag \\  
			&s.t.\ (\textrm{\ref{Con_edge_computing}}),(\textrm{\ref{Con_MA_range_0}}),(\textrm{\ref{Con_MA_distance_0}}).  \notag 
		\end{align}
	\end{subequations}

	It is computationally prohibitive to search the MA positioning directly, especially when the dimension of $\widetilde{\bm{r}}_0$ is large.    
	Thank to the fast convergence, easy implementation, and robust searching ability, the PSO algorithm is applied to search a sub-optimal of MA positioning \cite{xiao2023multiuser}. 
		
	We firstly randomly initialize $N$ particles with positions, i.e., $\mathcal{P}_0^{(0)}=\{\widetilde{\bm{r}}_0^{(0),1},\cdots,\widetilde{\bm{r}}_0^{(0),n},\cdots,\widetilde{\bm{r}}_0^{(0),N}\}$, where the position of each particle represents a candidate solution for the APV, i.e.,  
	\begin{equation*}
		\begin{aligned}
			\widetilde{\bm{r}}_0^{(0),n} 
			&=\big[\bm{r}_{0,1}^{(0),n},\cdots,\bm{r}_{0,m}^{(0),n},\cdots,\bm{r}_{0,M}^{(0),n}\big] \\
			&=\big[\underbrace{x_{0,1}^{(0),n},y_{0,1}^{(0),n}}_{\textrm{MA 1}},\cdots,
			   \underbrace{x_{0,m}^{(0),n},y_{0,m}^{(0),n}}_{\textrm{MA m}},\cdots,
			   \underbrace{x_{0,M}^{(0),n},y_{0,M}^{(0),n}}_{\textrm{MA M}}\big].
		\end{aligned}
	\end{equation*}	
	To characterize the movement of each particle, the velocities are randomly initialized as $\mathcal{V}_0^{(0)}=\{\bm{v}_0^{(0),1},\cdots,\bm{v}_0^{(0),n},\cdots,\bm{v}_0^{(0),N}\}$, where 
	\begin{equation*}
		\begin{aligned}
			\widetilde{\bm{v}}_0^{(0),n}
			&=[\bm{v}_{0,1}^{(0),n},\cdots,\bm{v}_{0,m}^{(0),n},\cdots,\bm{v}_{0,M}^{(0),n}] \\
			=&\big[\underbrace{v_{x,0,1}^{(0),n},v_{y,0,1}^{(0),n}}_{\textrm{MA 1}},\cdots, 	   	
			   \underbrace{v_{x,0,m}^{(0),n},v_{y,0,m}^{(0),n}}_{\textrm{MA m}}, \cdots,\underbrace{v_{x,0,M}^{(0),n},v_{y,0,M}^{(0),n}}_{\textrm{MA M}}\big].
		\end{aligned}
	\end{equation*}
	
	The key idea of the standard PSO algorithm is that each particle successively updates its position according to the personal best position found so far by itself, denoted by $\widetilde{\bm{r}}_{0,pbest}^{n}$, and the global best position found so far by all particles, denoted by $\widetilde{\bm{r}}_{0,gbest}$. However, the particles in the standard PSO algorithm are unable to explore the local neighborhoods to find a better solutions, which indicates that the standard PSO algorithm tends to be trapped in undesired sub-optima \cite{JunyingChen2005Partic}.  
	
	To cope with this problem, the variable local search is introduced to the standard PSO algorithm, shortened to PSO-VLS, to inspect all the personal best positions within the variable neighborhood. Specifically, $\widetilde{\bm{r}}_{0,gbest}$ is replaced by the best position found so far by particle $n$ and its neighbors, denoted by $\widetilde{\bm{r}}_{0,nbest}^{n}$. The neighborhood is determined by the distances between particles, calculated as $\Vert\widetilde{\bm{r}}_{0}^{(t),n}-\widetilde{\bm{r}}_{0}^{(t),\bar{n}} \Vert_2$, where $t$ denotes the iteration index. In addition, to improve the quality of $\widetilde{\bm{r}}_{0,nbest}^{n}$ with the increasing of iteration times, the number of particles within the neighborhood, denoted by $N^{(t),n}$, also increases. At the beginning, the neighborhood of each particle is set as the particle itself, i.e., $N^{(0),n}=1$. Then, the size of neighborhood grows in scale as iteration increases as follows:
	\begin{equation}  
		N^{(t+1),n}=\left\{
		\begin{aligned}
			&1, &&\textrm{if} \ t=0, \\
			&N^{(t),n} + \beta,   &&\textrm{if} \ N^{(t),n} + \beta < N, \\
			&N,   &&\textrm{if} \ N^{(t),n} + \beta\geq N, \\
		\end{aligned}
		\right.
	\end{equation}
	which indicates that the neighborhood will be gradually expanded to include all particles. 
	
	For each iteration, the velocity and position of particle $n$ are updated as 
	\begin{equation} \label{equation_velocity}
		\begin{aligned}
			\widetilde{\bm{v}}_0^{(t+1),n} = &\omega \widetilde{\bm{v}}_0^{(t),n} + 
			c_1 \phi_1 \big(\widetilde{\bm{r}}_{0,pbest}^{n}-\widetilde{\bm{r}}_0^{(t),n}\big)  \\
			&	+ c_2 \phi_2 \big(\widetilde{\bm{r}}_{0,nbest}^{n}-\widetilde{\bm{r}}_0^{(t),n}\big),   
		\end{aligned}
	\end{equation}
	and
	\begin{equation} \label{equation_position}
		\begin{aligned}
			\widetilde{\bm{r}}_{0}^{(t+1),n} = \widetilde{\bm{r}}_{0}^{(t),n} + \widetilde{\bm{v}}_0^{(t+1),n},
		\end{aligned}
	\end{equation}
	respectively.  Herein, $\omega\in[\omega_\text{min},\omega_\text{max}]$ is inertia weight of velocity, which linearly decreases with $t$ so as to improve the convergence speed, i.e.,  
	\begin{equation} \label{equation_omega}
		\omega = w_\textrm{max} - {(\omega_\textrm{max}-\omega_\textrm{min})\cdot(t/\mathcal{T})},
	\end{equation}
	where $\mathcal{T}$ is the default total number of iterations. $c_1$ and $c_2$ are the learning factors, indicating the step size of each particle moving towards the best position of itself and the best position of all particles, respectively. $\phi_1$ and $\phi_2$ are assigned randomly within $[0,1]$, whose aim is to enhance the randomness of the search for escaping from local optima.

	To avoid excessive roaming of particles, a velocity range, i.e., $[v_\textrm{min},v_\textrm{max}]$, is imposed to each entry of $\widetilde{\bm{v}}_0^{(t),n}$, i.e., 
	\begin{equation}  \label{equation_vx}
		v_{x,0,m}^{(t),n}=\left\{
		\begin{aligned}
			&v_\textrm{min},   &&\textrm{if} \ v_{x,0,m}^{(t),n}\leq v_\textrm{min}, \\
			&v_\textrm{max},   &&\textrm{if} \ v_{x,0,m}^{(t),n}\geq v_\textrm{max}, \\
			&v_{x,0,m}^{(t),n}, &&\textrm{otherwise},
		\end{aligned}
		\right.
	\end{equation}
	and
	\begin{equation}  \label{equation_vy}
		v_{y,0,m}^{(t),n}=\left\{
		\begin{aligned}
			&v_\textrm{max},   &&\textrm{if} \ v_{y,0,m}^{(t),n}\leq v_\textrm{min}, \\
			&v_\textrm{max},   &&\textrm{if} \ v_{y,0,m}^{(t),n}\geq v_\textrm{max}, \\
			&v_{y,0,m}^{(t),n}, &&\textrm{otherwise}.
		\end{aligned}
		\right.
	\end{equation}

	In addition, each entry of $\widetilde{\bm{r}}_0^{(t+1),n}$ should be checked to ensure that each particle stays inside the available region. As such, we have	
	\begin{equation}  \label{equation_x} 
		x_{0,m}^{(t),n}=\left\{
		\begin{aligned}
			&-A/2, 	&&\textrm{if} \ x_{0,m}^{(t),n}\leq -A/2, \\
			&A/2,  	&&\textrm{if} \  x_{0,m}^{(t),n}\geq A/2, \\
			&x_{0,m}^{(t),n}, &&\textrm{otherwise},
		\end{aligned}
		\right.
	\end{equation}
	and
	\begin{equation} \label{equation_y}
		y_{0,m}^{(t),n}=\left\{
		\begin{aligned}
			&-A/2, 	&&\textrm{if} \ y_{0,m}^{(t),n}\leq -A/2, \\
			&A/2,  	&&\textrm{if} \ y_{0,m}^{(t),n}\geq A/2,  \\
			&y_{0,m}^{(t),n}, &&\textrm{otherwise}.
		\end{aligned}
		\right.
	\end{equation}
	
	To evaluate the fitness of each particle, the fitness function is formulated as
	\begin{equation} \label{equation_fitness}
		\mathcal{F}_0(\widetilde{\bm{r}}_0^{(t),n})=\left\{
		\begin{aligned}
			&R(\widetilde{\bm{r}}_0^{(t),n}),	&&\textrm{if (\ref{Con_edge_computing}) and (\ref{Con_MA_distance_0}) hold}, \\
			&-1, &&\textrm{otherwise,}
		\end{aligned}
		\right.
	\end{equation}
	where  $R(\widetilde{\bm{r}}_0^{(t),n})=\sum_{k\in\mathcal{K}}\big(R_{L,k}(\widetilde{\bm{r}}_0^{(t),n})+R_{O,k}(\widetilde{\bm{r}}_0^{(t),n})\big)$.
	With the fitness evaluation conducted on the particles during each iteration, their personal and neighborhood best positions are updated until convergence or the total number of iterations is reached. The best position among the particles is selected to be a sub-optimal solution for $\widetilde{\bm{r}}_0$. 
	
	The details of the PSO-VLS algorithm for solving (PA2.2) are summarized in Algorithm \ref{Alg_PSO_VLS}, whose convergence is guaranteed by the following inequalities:
	\begin{equation*}
		\mathcal{F}_0(\widetilde{\bm{r}}_{0,pbest}^{(t+1),n}) \geq \mathcal{F}_0(\widetilde{\bm{r}}_{0,pbest}^{(t),n}),
	\end{equation*}
	and
	\begin{equation*}
		\mathcal{F}_0(\widetilde{\bm{r}}_{0,nbest}^{(t+1),n}) \geq \mathcal{F}_0(\widetilde{\bm{r}}_{0,nbest}^{(t),n}),
	\end{equation*}
	respectively. 
	Accordingly, we have 
	\begin{equation*}
		\mathcal{F}_0(\widetilde{\bm{r}}_{0}^{(t+1)}) \geq \mathcal{F}_0(\widetilde{\bm{r}}_{0}^{(t)}),
	\end{equation*}
	which indicates that fitness value of $\widetilde{\bm{r}}$ is non-decreasing during the iterations. In addition, the objective value of (PA2.2) is bounded due to (\ref{Con_energy_beamforming_matrix}). Therefore, the convergence of Algorithm \ref{Alg_PSO_VLS} is guaranteed. Compared to the standard PSO, each particle in PSO-VLS needs to calculate the distances to other particles to conform its neighbors. The computational complexity of Algorithm \ref{Alg_PSO_VLS} is $\mathcal{O}\left(\mathcal{T}N^2\right)$.	 
	
	\begin{algorithm}[ht] 
		\caption{PSO-VLS algorithm for solving (PA2.2).}
		\label{Alg_PSO_VLS}
		\begin{algorithmic}[1]
			\STATE Initialize $N$ particles with positions $\mathcal{P}_0^{(0)}$ and velocities $\mathcal{V}_0^{(0)}$, $\widetilde{\bm{r}}_{0,pbest}^{n}=\widetilde{\bm{r}}_{0}^{(0),n}$, and $\widetilde{\bm{r}}_{0,nbest}^{n}=\widetilde{\bm{r}}_{0}^{(0),n}$.
			\STATE Evaluate the fitness of each particle. 
			\FOR{$t=1$ to $\mathcal{T}$}
				\STATE Update the inertia weight $\omega$ according to (\ref{equation_omega}).
				\FOR{$n=1$ to $N$}
					\STATE Update the velocity of particle $n$ according to (\ref{equation_velocity}), (\ref{equation_vx}) and (\ref{equation_vy}).
					\STATE Update the position of particle $n$ according to  (\ref{equation_position}), (\ref{equation_x}) and (\ref{equation_y}).
					\STATE Evaluate the fitness of particle $n$ according to (\ref{equation_fitness}). 
					\IF{$\mathcal{F}_0(\widetilde{\bm{r}}_0^{(t),n})>\mathcal{F}_0(\widetilde{\bm{r}}_{0,pbest}^{n})$}
					\STATE $\widetilde{\bm{r}}_{0,pbest}^{n} \leftarrow \widetilde{\bm{r}}_0^{(t),n}$. 
					\ENDIF
					\IF{$\max\{\textrm{neighborhood fitness values}\}>\mathcal{F}_0(\widetilde{\bm{r}}_{0,nbest}^{n})$}
					\STATE $\widetilde{\bm{r}}_{0,nbest}^{n}=\arg\max\{\textrm{neighborhood fitness values}\}$.
					\ENDIF  
				\ENDFOR
			\ENDFOR 
			\STATE Obtain $\widetilde{\bm{r}}_{0}=\arg\max\big\{\mathcal{F}_0(\widetilde{\bm{r}}_{0,nbest}^{n}),\ n=1,\cdots,N\big\}$.
		\end{algorithmic}
	\end{algorithm}

	\subsubsection{Optimizing $\mathbf{Q}$} Given $\widetilde{\bm{r}}_0$, (PA2.1) is reduced to
	\begin{subequations} 
		\begin{align}
			(\textrm{PA2.3}):\ &\max_{\mathbf{Q}\succeq0,\bm{z}}\sum_{k\in\mathcal{K}}\big(R_{L,k}(\mathbf{Q})+R_{O,k}(\mathbf{Q})\big) \notag \\  
			&\qquad s.t.\ (\textrm{\ref{Con_edge_computing}}), (\textrm{\ref{Con_energy_beamforming_matrix}}), \notag \\
			z_k =&  \textrm{exp}\left(-a_k\Big(\bm{h}_k^T(\widetilde{\bm{r}}_0)\mathbf{Q}\left(\bm{h}_k^T(\widetilde{\bm{r}}_0)\right)^H-b_k\Big)\right), \label{Con_z_k}
		\end{align}
	\end{subequations}
	where $z_k$ is the auxiliary variable, and $\bm{z}=\{z_k,k\in\mathcal{K}\}$. Then, $\Xi_k$ can be rewritten as $\Xi_k= \dfrac{X_k}{1+z_k}-Y_k$. Since  $\Xi_k$ is non-convex of $z_k$, the SCA method is applied to approximate $\Xi_k$ as
	\begin{equation*}
		\Xi_k=X_k\left(\dfrac{1}{1+z_k^{(r)}} - \dfrac{z_k-z_k^{(r)}}{\big(1+z_k^{(r)}\big)^2}\right)-Y_k,
	\end{equation*}
	where $z_k^{(r)}$ is a feasible solution in the $r$-th iteration. 
	Accordingly, (PA2.3) is simplified to a convex problem. 
	
	\subsection{Optimization of $\{\mathbf{w},\widetilde{\bm{r}}_k,k\in\mathcal{K}\}$}
	Given $\{\bm{\tau},\bm{p},\bm{f}\}$ and $\{\widetilde{\bm{r}}_0, \mathbf{Q}\}$, (PA) is reduced to 
	\begin{subequations} 
		\begin{align}
			(\textrm{PA3}):\ &\max_{\mathbf{w},\widetilde{\bm{r}}_k,k\in\mathcal{K}}\sum_{k\in\mathcal{K}}(R_{L,k}+R_{O,k})  \\ 
			\quad s.t.\ 
			& \textrm{(\ref{Con_edge_computing})}, \notag \\
			& \bm{r}_{k,m}\in\mathcal{C},\ m\in\mathcal{M}, k\in\mathcal{K},  \label{Con_MA_range_offload} \\
			 \Vert\bm{r}_{k,m}-\bm{r}_{k,n}\Vert_2& \geq D,\ 1\leq m\neq n\leq M, k\in\mathcal{K}. \label{Con_MA_distance_offload}
		\end{align}
	\end{subequations}

	Before solving (PA3), it is commonly believed that (\ref{Con_edge_computing}) can be omitted safely, due to the fact that a better SCR performance can be obtained by maximizing the capacity. Accordingly, removing the terms irrelevant to $\{\widetilde{\bm{r}}_k,\bm{w}_k,k\in\mathcal{K}\}$, (PA3) can be split into $K$ independent sub-problems, i.e.,
	\begin{subequations} 
		\begin{align}
			(\textrm{PA3.1}):\ &\max_{\bm{w}_k,\widetilde{\bm{r}}_k}R_{O,k}  \\ 
			\quad s.t.\ 
			& (\textrm{\ref{Con_MA_range_offload}}),(\textrm{\ref{Con_MA_distance_offload}}). \notag
		\end{align}
	\end{subequations}

	Given $\widetilde{\bm{r}}_k$, the optimal $\bm{w}_k$ can be designed by the MRC criterion \cite{ChenPengcheng2022MultiIRS}, i.e., 
	\begin{equation} \label{equation_w} 
		\bm{w}_{k} ={\sqrt{p_k}\bm{h}_{k}(\widetilde{\bm{r}}_k)}/{\sigma_0^2}, \ k\in\mathcal{K}.
	\end{equation} 
	Then, (PA3.1) can be equivalently  to the channel gain maximization problem as follows
	\begin{subequations} 
		\begin{align}
			(\textrm{PA3.2}):\ &\max_{\widetilde{\bm{r}}_k}, \Vert\bm{h}_k(\widetilde{\bm{r}}_k)\Vert^2 \notag \\ 
			&s.t.\ 
			 (\textrm{\ref{Con_MA_range_offload}}), (\textrm{\ref{Con_MA_distance_offload}}). \notag 
		\end{align}
	\end{subequations}

	The PSO-VLS algorithm is also applied to solve (PA3.2), and the fitness of  particle $n$ during the $t$-th iteration is evaluated by  
	\begin{equation*}
		\mathcal{F}_k(\widetilde{\bm{r}}_k^{(t),n})=\left\{
		\begin{aligned}
			&\Vert\bm{h}_k(\widetilde{\bm{r}}_k^{(t),n})\Vert^2, &&\textrm{if (\ref{Con_MA_distance_offload}) holds}, \\
			&-1, &&\textrm{otherwise.}
		\end{aligned}
		\right.
	\end{equation*}

	\subsection{Convergence and Complexity Analysis}
	\begin{algorithm}[t] 
		\caption{AO framework for solving (PA).}
		\label{Alg_AO_dynamic_MA}
		\begin{algorithmic}[1]
			\STATE Initialize $\mathbf{w}$, $\widetilde{\bm{r}}_0$, $\mathbf{Q}$ and $\bm{z}$.
			\FOR{$k=1$ to $K$}
				\STATE Recover $\widetilde{\bm{r}}_k$ by solving (PA3.2) with the PSO-VLS algorithm.
			\ENDFOR
			\REPEAT 
			\STATE Update $\bm{\tau}$ and $\bm{e}$ by solving (PA1.1) with CVX \cite{grant2008cvx}, and calculate $p_k=e_k/\tau_k$, $\forall k\in\mathcal{K}$. \label{step_tau_e}
			\STATE Calculate $\beta_k=p_k\tau_k/{\tau_0\Xi_k}$, $\forall k\in\mathcal{K}$.
			\STATE Update $\widetilde{\bm{r}}_0$ by solving (PA2.2) with the PSO-VLS algorithm shown in Algorithm \ref{Alg_PSO_VLS}.
			\REPEAT 
			\STATE Update $\mathbf{Q}$ and $\bm{z}$ by CVX \cite{grant2008cvx}.
			\UNTIL{Convergence.}
			\UNTIL{Convergence.}
			\STATE Calculate $\bm{w}_k$ according to (\ref{equation_w}), $\forall k\in\mathcal{K}$. 
			\RETURN $\bm{\tau}$, $\bm{p}$, $\bm{f}$, $\mathbf{w}$, $\widetilde{\bm{r}}$, $\mathbf{Q}$.
		\end{algorithmic}
	\end{algorithm}
	The procedure for solving (PA) is summarized in Algorithm \ref{Alg_AO_dynamic_MA}. As mentioned above, the PSO-VLS algorithm for solving (PA2.2) is guaranteed to converge. Similarly, the channel gain of each WD for task offloading increases by optimizing $\widetilde{\bm{r}}_k$ with PSO-VLS algorithm, i.e., $\Vert\bm{h}_k({\widetilde{\bm{r}}_k^{(t+1),n}})\Vert^2\geq\Vert\bm{h}_k({\widetilde{\bm{r}}_k^{(t),n}})\Vert^2$. More tasks can be executed by setting $\{\bm{\tau},\bm{p},\bm{f}\}$ more properly with the guidance of other variables. In addition, the SCR performance increases by optimizing $\mathbf{Q}$ with SCA method, because more energy can be harvested for task execution. Meanwhile, the performance of the MA-enhanced WP-MEC system is bounded by the communication resource and computing capability. Therefore, the convergence of Algorithm \ref{Alg_AO_dynamic_MA} is guaranteed.

	The complexity is then analyzed as follows. (PA1.1) can be solved with complexity of $\mathcal{O}\big((2K+2)^{3.5}\big)$, where $(2K+2)$ is the number of variables. In addition, $\mathbf{Q}$ and $\bm{z}$ can be obtained by solving (PA2.3) with worst-case complexity of $\mathcal{O}(L_\textrm{SCA}(M^2+K)^{3.5})$ \cite{LuoZhiquan2010Semidefinite}, where $M^2$ is the number of variables in matrix $\mathbf{Q}$, and $L_\textrm{SCA}$ is the number of iterations required to converge. Therefore, the total complexity of Algorithm \ref{Alg_AO_dynamic_MA} is $\mathcal{O}\Big(K\mathcal{T}N^2 + L_\textrm{AO}\left((2K+2)^{3.5} + L_\textrm{SCA}(M^2+K)^{3.5} + \mathcal{T}N^2 \right)\Big)$, where $L_\textrm{AO}$ is the number of iterations required till convergence.

	\section{Extension to Semi-Dynamic and Static MA Positioning}
	\label{section_solution_for_semi-dynamic_MA_positioning and Static MA Positioning}
	In this section, the AO framework with the PSO-VLS algorithm is extended to solve the formulated problems for the other two types of MA positioning configurations, i.e., semi-dynamic MA positioning and static MA positioning. It is observed that the variables, i.e.,  $\{\bm{\tau}$, $\mathbf{Q}$, $\mathbf{w}$, $\bm{p}$, $\bm{f}\}$, in (PB) and (PC) can be optimized using the similar methods as (PA). As such, we only focus on the optimization of APVs.
	
	\subsection{Optimization of $\widetilde{\bm{r}}_u$ for Semi-dynamic MA Positioning}
	In the case, the APV optimization sub-problem for (PB) is 
	\begin{subequations} 
		\begin{align}
			(\textrm{PB1}):\ &\max_{\widetilde{\bm{r}}_u}\sum_{k\in\mathcal{K}}R_{O,k}(\widetilde{\bm{r}}_u) \notag\\ 
			s.t.&\ 
			(\textrm{\ref{Con_edge_computing}}),(\textrm{\ref{Con_MA_range_semi_dynamic}}), (\textrm{\ref{Con_MA_distance_semi_dynamic}}), \notag
		\end{align}
	\end{subequations}
	where
	\begin{equation*}
		R_{O,k}(\widetilde{\bm{r}}_u) = \dfrac{B\tau_k}{T}\log_2\Big(1+\dfrac{\beta_k\tau_{0}\Xi_k|\bm{w}_k^H\bm{h}_k(\widetilde{\bm{r}}_u)|^2}{\tau_{k}\Vert\bm{w}_{k}^H\Vert^2\sigma_0^2}\Big).
	\end{equation*}
	
	The PSO-VLS algorithm is applied to solve (PB1), and the fitness of particle $n$ in the $t$-th iteration can be evaluated as  
	\begin{equation*}
		\mathcal{F}_u(\widetilde{\bm{r}}_u^{(t),n})=\left\{
		\begin{aligned}
			&R_{O,k}(\widetilde{\bm{r}}_u^{(t),n}),	&&\textrm{if (\ref{Con_edge_computing}) and (\ref{Con_MA_distance_semi_dynamic}) hold}, \\
			&-1, &&\textrm{otherwise.}
		\end{aligned}
		\right.
	\end{equation*}

	The procedure for solving (PB) is summarized in Algorithm \ref{Alg_AO_semi_dynamic_MA}, whose convergence can be demonstrated analogous to that of Algorithm  \ref{Alg_AO_dynamic_MA}. Based on the analyses above, the total complexity of Algorithm \ref{Alg_AO_semi_dynamic_MA} is $\mathcal{O}\Big(L_\textrm{AO}\left((2K+2)^{3.5} + L_\textrm{SCA}(M^2+K)^{3.5} + 2\mathcal{T}N^2 \right)\Big)$.
	\begin{algorithm}[ht] 
		\caption{AO framework for solving (PB).}
		\label{Alg_AO_semi_dynamic_MA}
		\begin{algorithmic}[1]
			\STATE Initialize $\mathbf{w}$, $\widetilde{\bm{r}}_0$, $\widetilde{\bm{r}}_u$, $\mathbf{Q}$ and $\bm{z}$.
			\REPEAT 
			\STATE Update time allocation $\bm{\tau}$ and $\bm{e}$ by solving (PA1.1) with CVX \cite{grant2008cvx}, and calculate $p_k=e_k/\tau_k$, $k\in\mathcal{K}$.
			\STATE Calculate $\beta_k=p_k\tau_k/{\tau_0\Xi_k}$, $k\in\mathcal{K}$.
			\STATE Update $\widetilde{\bm{r}}_0$ by using the PSO-VLS algorithm.
			\REPEAT 
			\STATE Update $\mathbf{Q}$ and $\bm{z}$ by CVX \cite{grant2008cvx}.
			\UNTIL{Convergence.}
			\STATE Recover $\widetilde{\bm{r}}_u$ by solving (PB1) with the PSO-VLS algorithm.
			\UNTIL{Convergence.}
			\STATE Calculate $\bm{w}_k$ according to (\ref{equation_w}), $k\in\mathcal{K}$. 
			\RETURN $\bm{\tau}$, $\bm{p}$, $\bm{f}$, $\mathbf{w}$, $\widetilde{\bm{r}}_0$, $\widetilde{\bm{r}}_u$, $\mathbf{Q}$.
		\end{algorithmic}
	\end{algorithm}

	\subsection{Optimization of $\widetilde{\bm{r}}_s$ for Static MA Positioning}	
	In the case, the APV optimization sub-problem for (PC) is
	\begin{subequations} 
		\begin{align}
			(\textrm{PC1}):\ \max_{\widetilde{\bm{r}}_s}&\sum_{k\in\mathcal{K}}\big(R_{L,k}(\widetilde{\bm{r}}_s)+R_{O,k}(\widetilde{\bm{r}}_s)\big) \notag \\ 
			& s.t.\ (\textrm{\ref{Con_edge_computing}}), (\textrm{\ref{Con_MA_range_static}}),
			(\textrm{\ref{Con_MA_distance_static}}). \notag
		\end{align}
	\end{subequations}
	where
	\begin{equation*} 
		R_{L,k}(\widetilde{\bm{r}}_s) = \dfrac{1}{\varphi_k}\Big(\dfrac{(1-\beta_k)\tau_{0}\Xi_k(\widetilde{\bm{r}}_s)}{\kappa T}\Big)^{\tfrac{1}{3}}, 
	\end{equation*}
	and
	\begin{equation*} 
		R_{O,k}(\widetilde{\bm{r}}_s) = \dfrac{B\tau_k}{T}\log_2\Big(1+\dfrac{\beta_k\tau_{0}\Xi_k(\widetilde{\bm{r}}_s)|\bm{w}_k^H\bm{h}_k(\widetilde{\bm{r}}_s)|^2}{\tau_{k}\Vert\bm{w}_{k}^H\Vert^2\sigma_0^2}\Big).
	\end{equation*}
	
	We solve (PC1) by the PSO-VLS algorithm and calculated the fitness of particle $n$ in the $t$-th iteration as  
	\begin{equation*}
		\mathcal{F}_s(\widetilde{\bm{r}}_s^{(t),n})=\left\{
		\begin{aligned}
			&R(\widetilde{\bm{r}}_s^{(t),n}),	&&\textrm{if (\ref{Con_edge_computing}) and (\ref{Con_MA_distance_static}) hold}, \\
			&-1, &&\textrm{otherwise,}
		\end{aligned}
		\right.
	\end{equation*}
	where $R(\widetilde{\bm{r}}_s^{(t),n})=\sum_{k\in\mathcal{K}}\big(R_{L,k}(\widetilde{\bm{r}}_s^{(t),n})+R_{O,k}(\widetilde{\bm{r}}_s^{(t),n})\big)$. 
	
	The procedure for solving (PC) is summarized in Algorithm \ref{Alg_AO_static_MA}. The convergence of Algorithm \ref{Alg_AO_static_MA} can be verified in a similar way as Algorithm \ref{Alg_AO_dynamic_MA}. The complexity of Algorithm \ref{Alg_AO_static_MA} is $\mathcal{O}\Big(L_\textrm{AO}\left((2K+2)^{3.5} + L_\textrm{SCA}(M^2+K)^{3.5} + \mathcal{T}N^2 \right)\Big)$.	
	\begin{algorithm}[h] 
		\caption{AO framework for solving (PC).}
		\label{Alg_AO_static_MA}
		\begin{algorithmic}[1]
			\STATE Initialize $\mathbf{w}$, $\widetilde{\bm{r}}_s$, $\mathbf{Q}$ and $\bm{z}$.
			\REPEAT 
				\STATE Update time allocation $\bm{\tau}$ and $\bm{e}$ by solving (PA1.1) with CVX \cite{grant2008cvx}, and calculate $p_k=e_k/\tau_k$, $k\in\mathcal{K}$.
				\STATE Calculate $\beta_k=p_k\tau_k/{\tau_0\Xi_k}$, $k\in\mathcal{K}$.
				\STATE Update $\widetilde{\bm{r}}_s$ by solving (PC1) with the PSO-VLS algorithm.
				\REPEAT 
				\STATE Update $\mathbf{Q}$ and $\bm{z}$ by CVX \cite{grant2008cvx}.
				\UNTIL{Convergence.}
			\UNTIL{Convergence.}
			\STATE Calculate $\bm{w}_k$ according to (\ref{equation_w}), $k\in\mathcal{K}$. 
			\RETURN $\bm{\tau}$, $\bm{p}$, $\bm{f}$, $\mathbf{w}$, $\widetilde{\bm{r}}_s$, $\mathbf{Q}$.
		\end{algorithmic}
	\end{algorithm}

	\section{Numerical Results}
	\label{section_numerical_results}
	Numerical results are provided to evaluate the SCR performance of the MA-enhanced WP-MEC systems, where the available region is set as a square area centered at $(0.0)$ meter (m), i.e., $\mathcal{C}=[-A/2,A/2]\times[-A/2,A/2]$. The WDs are randomly located around the HAP with distances following uniform distribution from $7$ m to $8$ m \cite{ChenPengcheng2023ComputationalRate}. It is assumed that each entry in PRV $\bm{g}_k$ follows $g_{k,l}\sim\mathcal{CN}(0,C_0d_k^{-\alpha})$, where $C_0$ denotes the large-scale fading at reference distance $1$ m, and $\alpha$ is the path-loss exponent. The other parameters are listed in TABLE \ref{Table_parameter_settings}.
	\begin{table} [t]
		\centering
		\caption{Parameter Settings}
		\label{Table_parameter_settings}
		\begin{tabular}{|c|c|c|}
			\hline
			\textbf{Description} 			   & \textbf{Parameter and Value} \\ \hline
			\multirow{6}*{Communication model} & $T=1$ second, $M=8$, $K=6$ \\ \cline{2-2}
											   & $B=50$ KHz	\\ \cline{2-2}
											   & $\lambda=0.1$ m, $L_k=10$,  \cite{xiao2023multiuser} \\ \cline{2-2}
											   & $A=3\lambda$, $D=0.5\lambda$ \cite{xiao2023multiuser} \\ \cline{2-2}
											   & $C_0=(\lambda/4\pi)^2$, $\alpha=2.2$ \\ \cline{2-2}
											   & $P_\textrm{max}=40$ dBm, $\sigma_0^2=-80$ dBm \\  \hline
			\multirow{1}*{EH model \cite{MaoSun2022Reconfigurable}} & $M_k=0.024$ W, $a_k=150$, $b_k=0.014$	\\ \hline
			\multirow{3}*{Computation model\cite{ChenPengcheng2023ComputationalRate}}&$f_E=0.4$ GHz  \\ \cline{2-2} 
											   & $\varphi_k=1000$ cycles/bit \\ \cline{2-2}
											   & $\kappa=10^{-26}$ Watt/Hz$^3$	\\ \hline
			\multirow{4}*{PSO-VLS algorithm\cite{xiao2023multiuser}}& $\mathcal{T} = 200$, $\beta=1$ \\ \cline{2-2}
											   & $\omega_\textrm{min}=0.4$, $\omega_\textrm{max}=0.9$\\ \cline{2-2}
											   & $c_1=1.4$, $c_2 = 1.4$ \\ \cline{2-2}
											   & $v_\textrm{min}=-0.5\lambda$, $v_\textrm{max}=0.5\lambda$ \\ \hline
		\end{tabular}
	\end{table}
	
	\subsection{Convergence Behaviors}
\begin{figure}[t]
		\centering\includegraphics[width=9.2cm]{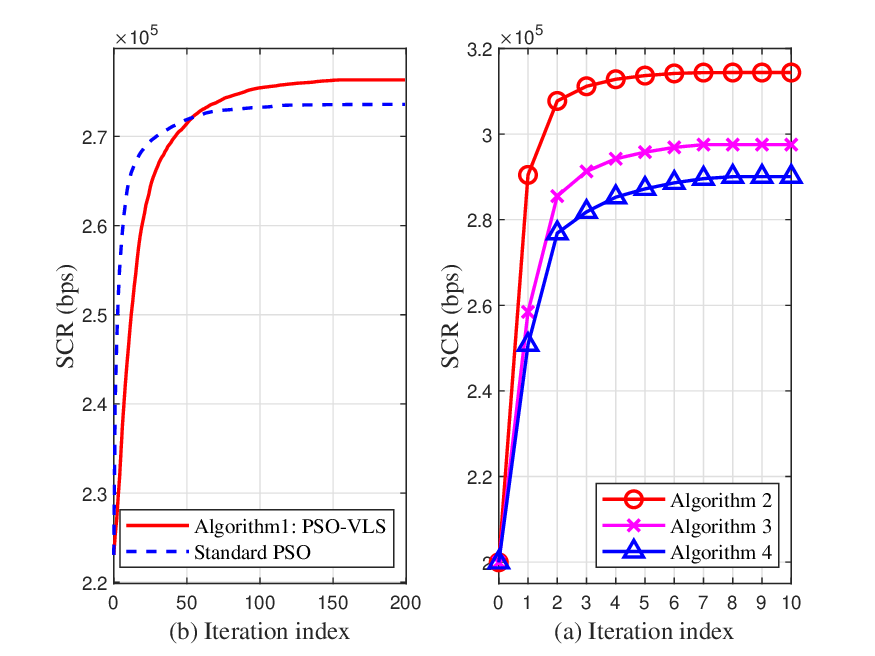}
		\caption{Convergence behaviors of the proposed algorithms.}
		\label{S_convergence}
	\end{figure}

	Fig.~\ref{S_convergence} depicts the convergence behaviors of the PSO-VLS algorithm (i.e., Algorithm \ref{Alg_PSO_VLS}) and the AO frameworks for the three cases with different types of MA positioning configurations (i.e., Algorithms \ref{Alg_AO_dynamic_MA}$\sim$\ref{Alg_AO_static_MA}). From Fig.~\ref{S_convergence}(a), it is obvious that the PSO-VLS algorithm achieves a better SCR performance than the standard PSO algorithm after a slightly more iterations. The reason is that each particle in the PSO-VLS algorithm is reasonable for the search of its variable neighborhoods instead of all the particles. The diversity of particles in the PSO-VLS algorithm keep longer, and slightly more iterations are required. Fig.~\ref{S_convergence}(b) shows that the SCRs obtained by Algorithms \ref{Alg_AO_dynamic_MA}$\sim$\ref{Alg_AO_static_MA} increase with iteration index and converge within almost the same number of iterations, which demonstrates the fast convergence and good robustness of the proposed AO frameworks. Specifically, the SCR obtained by Algorithm \ref{Alg_AO_dynamic_MA} increases from $2.0\times10^{5}$ bps to $3.144\times10^{5}$ bps, which yields $57.2\%$ performance improvement. In addition, there are $48.7\%$ and $45.1\%$ performance improvements for Algorithm \ref{Alg_AO_semi_dynamic_MA} and Algorithm \ref{Alg_AO_static_MA}, respectively. Algorithm \ref{Alg_AO_dynamic_MA} achieves the highest performance improvement over Algorithms \ref{Alg_AO_semi_dynamic_MA} and \ref{Alg_AO_static_MA}, because more spatial DoFs can be obtained by flexibly adjusting the positions of MAs during each time slot, inevitably incurring a certain magnify in computational complexity.

	\subsection{Effectiveness of the AO Frameworks}
	\begin{figure}[t]
		\centering\includegraphics[width=9.2cm]{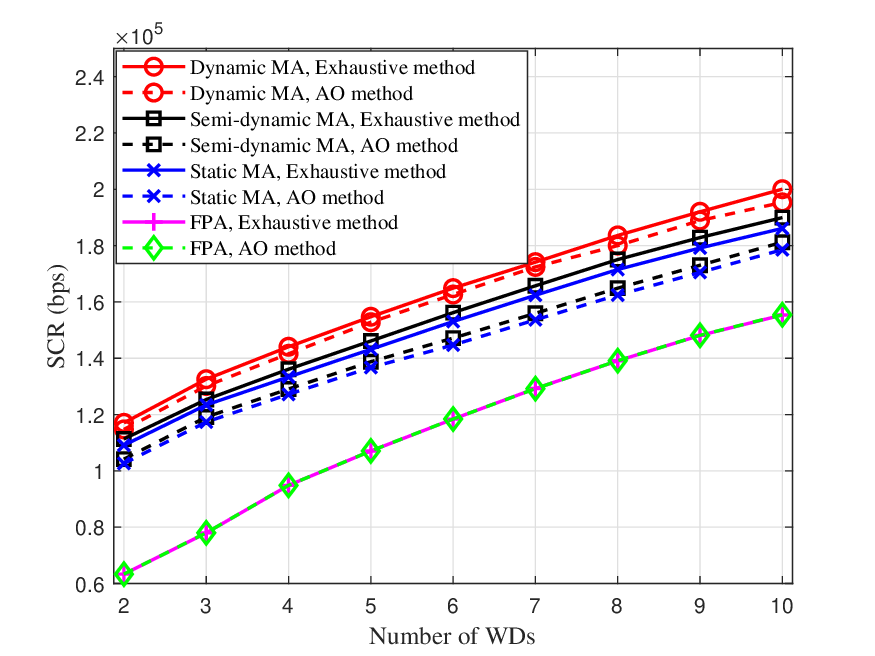}
		\caption{The performance comparisons between the proposed AO frameworks and the exhaustive method with $M=2$.}
		\label{S_exhaustive}
	\end{figure}
	To evaluate the effectiveness of the proposed AO frameworks, the performance comparisons between the obtained solutions and the optimal solutions achieved by the exhaustive method are carried out. Before solving the formulated problems (i.e., PA, PB, and PC) by the exhaustive method, some conversions are made. Firstly, the continuous spatial region $\mathcal{C}$ is equally divided into 16 parts, whose centers can be used for antenna deployments.Then, to further reduce the computational complexity of the exhaustive method, the number of antennas is set to 2 (i.e., $M=2$). At last, the continuous value of $\tau_0$ is  assumed to take a finite number of values, i.e., obtained by equally quantizing the interval $[0,T]$ into $100$ levels. Thus, we have $\tau_0\in\{0,\Delta T,2\Delta T,\cdots,T\}$, where $\Delta T = T/100$. Based on these conversions, the formulated problems are resolved by the AO frameworks and the exhaustive method.  

	Fig.~4 depicts the performance comparisons between the proposed AO frameworks and the exhaustive method. Compared to the exhaustive method, the AO frameworks achieve near-to-optimal performances with only 1.68 \%, 5.42 \% and 4.96 \% degradations in average for the schemes with dynamic MA positioning, semi-dynamic MA positioning, and static MA positioning, respectively. It demonstrates the effectiveness of the AO frameworks. It is worth noting that the AO frameworks achieve almost the same performance with the exhaustive method for the FPA scheme, which further verifies the effectiveness of the AO frameworks.

	\subsection{Performance Comparison}
	The following benchmark schemes are defined for performance comparison:
	\begin{itemize}
		\item \textbf{FPA}: The HAP is equipped with FPA-based uniform planar array (UPA) with $M$ antennas, spaced by $D$ \cite{xiao2023multiuser}. Specifically, each antenna at the HAP has a fixed position. The corresponding energy beamforming matrix (i.e., $\mathbf{Q}$) can be solved via the SCA method shown in Section III B 2). In addition, the receive combing vectors (i.e., $\mathbf{w}$) can be calculated by using (\ref{equation_w}).

		\item \textbf{PSO}: The APV of all MAs are optimized by the standard PSO algorithm \cite{xiao2023multiuser}.  Specifically, compared to the PSO-VLS algorithm, the best position found so far by the $n$-th particle and its neighbors (i.e., $\widetilde{\bm{r}}_{0,nbest}^{n}$) is replaced by the global best position found so far by all particles (i.e., $\widetilde{\bm{r}}_{0,gbest}$). For each iteration, the velocity and position of the $n$-th particle are updated as $\widetilde{\bm{v}}_0^{(t+1),n} = \omega \widetilde{\bm{v}}_0^{(t),n} + c_1 \phi_1 \big(\widetilde{\bm{r}}_{0,pbest}^{n}-\widetilde{\bm{r}}_0^{(t),n}\big) + c_2 \phi_2 \big(\widetilde{\bm{r}}_{0,gbest}-\widetilde{\bm{r}}_0^{(t),n}\big)$ and (\ref{equation_position}), respectively.
		
		\item \textbf{Offloading-only}: The harvested energy of each WD is exhausted to perform task offloading, instead of implementing the local computing. That is to say, the local computing is in unavailable for this benchmark scheme.
	\end{itemize}

	\begin{figure}[t]
		\centering\includegraphics[width=9.2cm]{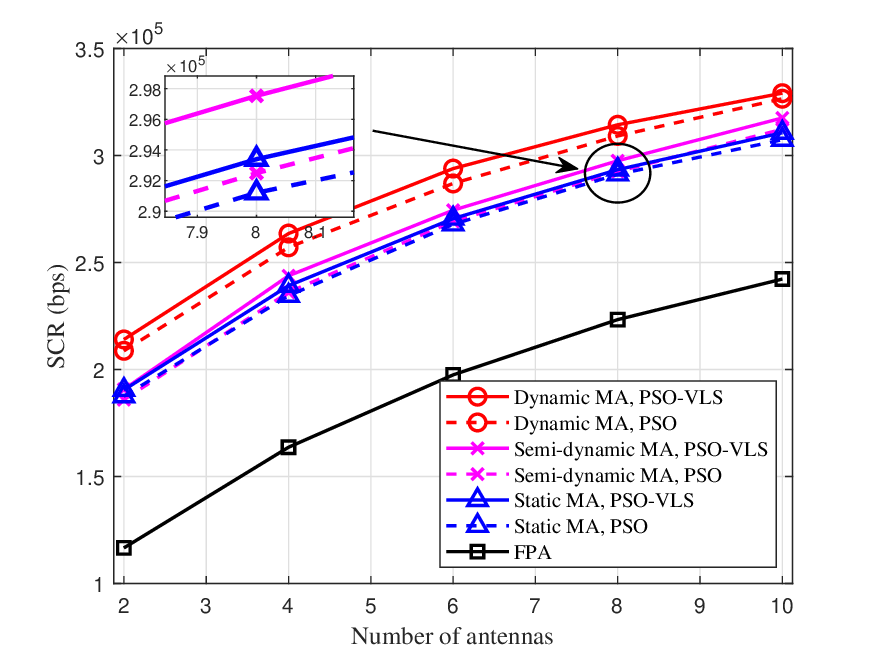}
		\caption{SCR versus the number of antennas at the HAP.}
		\label{S_num_of_antenna}
	\end{figure}
	\begin{figure}[t]
		\centering\includegraphics[width=9.2cm]{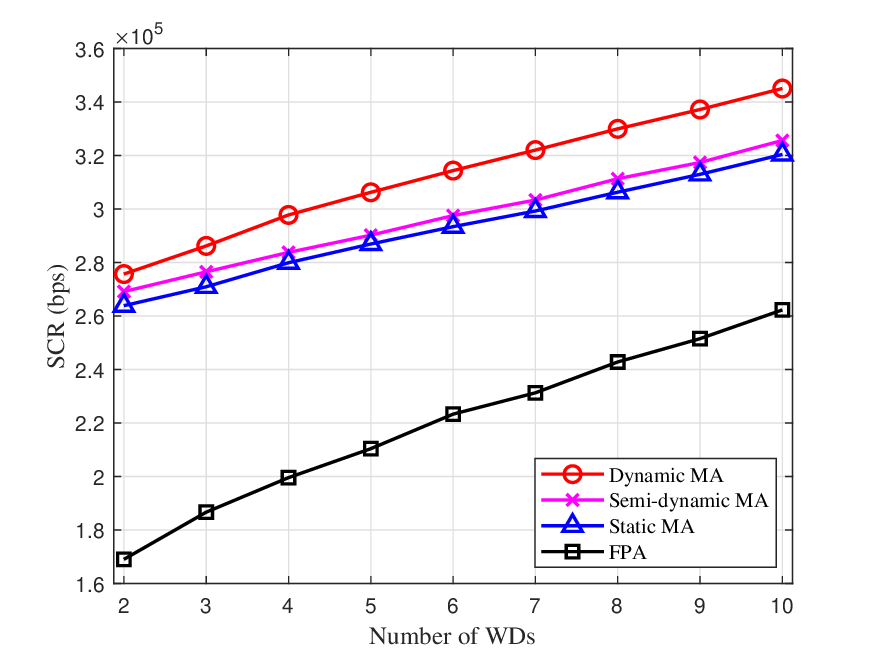}
		\caption{SCR versus the number of WDs.}
		\label{S_num_of_WD}
	\end{figure}
	\begin{figure}[t]
		\centering\includegraphics[width=9.2cm]{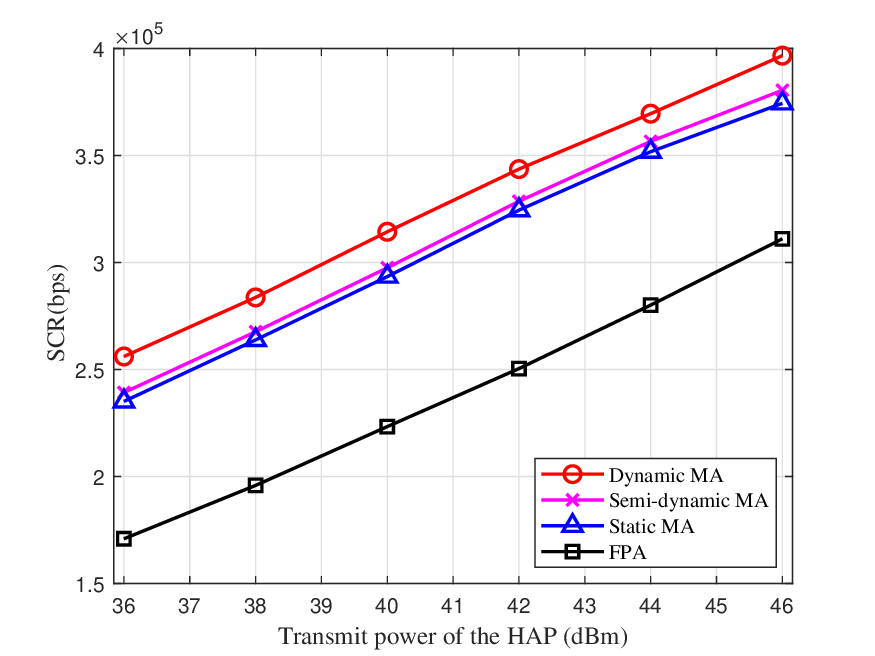}
		\caption{SCR versus the power budget of HAP.}
		\label{S_HAP_power}
	\end{figure}
	\begin{figure}[t]
		\centering\includegraphics[width=9.2cm]{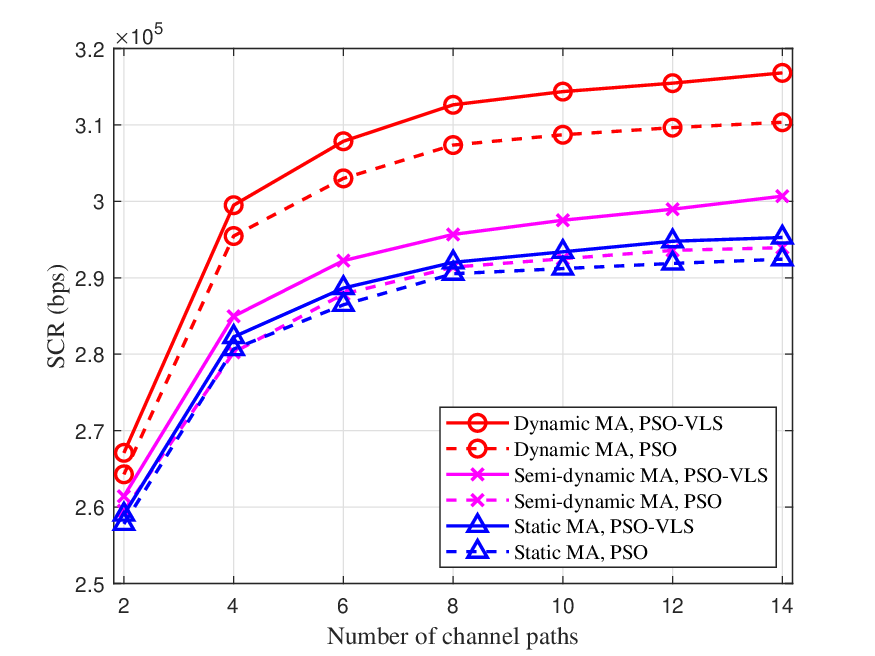}
		\caption{SCR versus the number of channel paths.}
		\label{S_num_of_path}
	\end{figure}
	\begin{figure}[t]
		\centering\includegraphics[width=9.2cm]{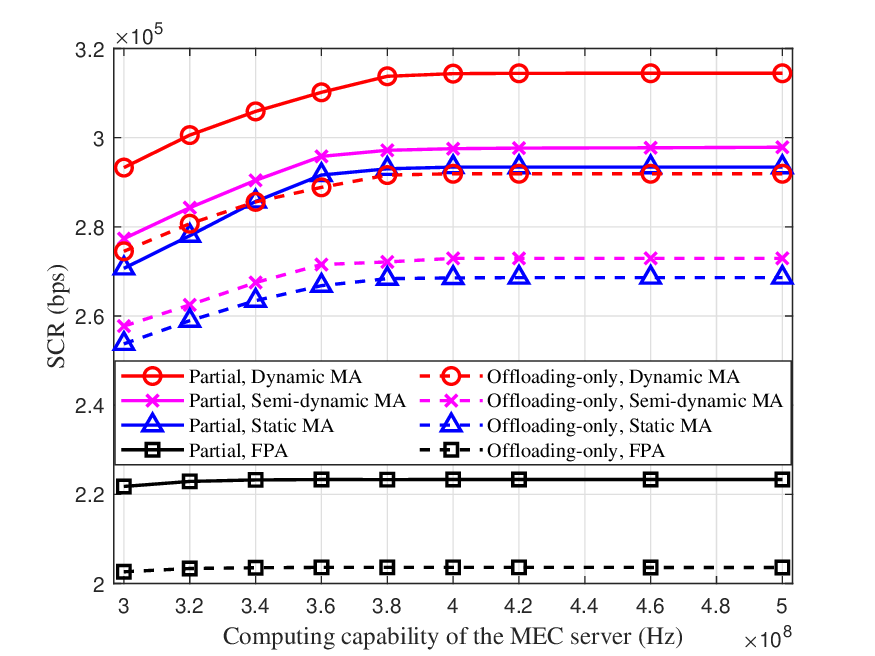}
		\caption{SCR versus the edge computing capability.}
		\label{S_capability}
	\end{figure}
	\begin{figure}[t]
		\centering\includegraphics[width=9.2cm]{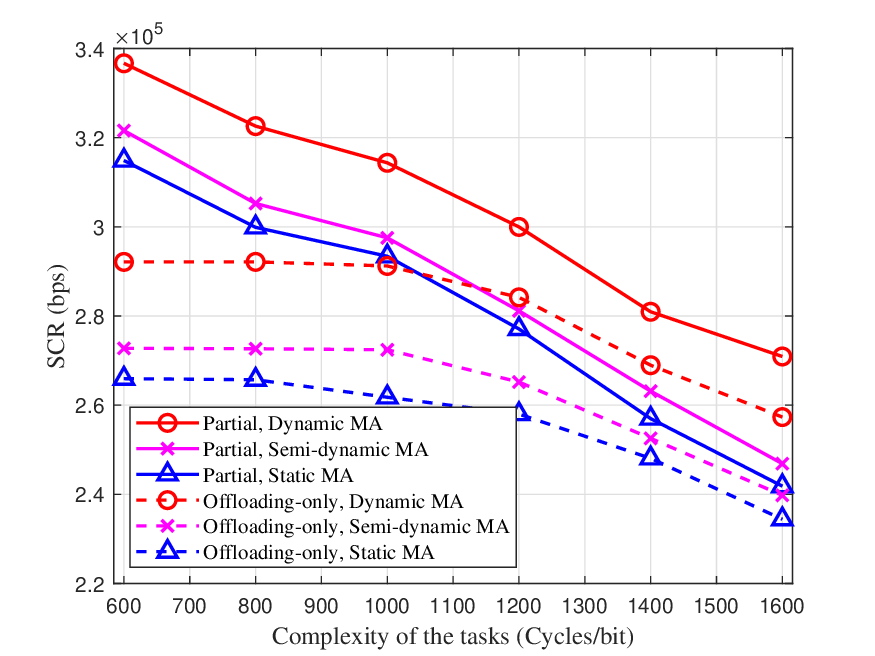}
		\caption{SCR versus the complexity of the tasks.}
		\label{S_complexity}
	\end{figure}
	Fig.~\ref{S_num_of_antenna} depicts the SCR versus the number of MAs, i.e., $M$. It can be seen that the SCRs of all schemes increase with $K$, which can be explained as follows. On the one hand, the HAP can perform a more fine-grained energy emission during the WPT stage by taking into account the spatial diversities of all WDs. On the other hand, the stronger beamforming gain can be obtained for task offloading. The MA schemes significantly outperform the FPA scheme, because the MA schemes obtain more spatial DoFs by properly optimizing the positions of MAs. Obviously, the  PSO-VLS schemes obtain better performance than that of the counterpart PSO schemes, which further validates the benefit of PSO-VLS algorithm in escape from the local optima. 
	
	Fig.~\ref{S_num_of_WD} shows the SCR versus the number of WDs, i.e., $K$. It is observed that the SCR performance increases with $K$ for all schemes. The MA schemes outperform the FPA scheme, which again demonstrates the superiority of MAs over the FPAs in the WP-MEC systems. Specifically, the schemes with dynamic, semi-dynamic, and static MA positioning achieve about $42.4\%$, $35.3\%$ and $33.2\%$ SCR gains over the FPA scheme, respectively. In addition, as $K$ increases, the performance gaps between the dynamic MA scheme and the other two MA schemes increase. It is due to the fact that the APV for each WD in the dynamic MA scheme is optimized for easy task offloading.
	
	Fig.~\ref{S_HAP_power} depicts the SCRs of different schemes versus the power budget of HAP, i.e., $P_\textrm{max}$. As expected, with the increase of $P_\textrm{max}$, the SCRs of all schemes increase. The MA schemes achieve more performance gains over the FPA scheme, which again demonstrates the superiority of MA-enhanced WP-MEC over the conventional  WP-MEC with FPAs. Significantly, the MA schemes achieve about $42.5\%$ and $23.4\%$ performance gains over the FPA scheme when $P_\textrm{max}=36$ dBm and  $P_\textrm{max}=46$ dBm, respectively. This indicates that the MA schemes are more applicable for the WP-MEC systems with low power budget over the FPA scheme. 	
	
	Fig.~\ref{S_num_of_path} illustrates the SCR versus the number of channel paths of each WD, i.e., $L_k$. The SCRs of all schemes increase with $L_k$ because higher multi-path diversity can be exploited and more design flexibility of energy beams and receive combing vectors at the HAP can be obtained. In addition, the performance gaps between the PSO-VLS schemes and their counterpart PSO schemes increases with $L_k$, which further validates the superiority of PSO-VLS algorithm over the standard PSO algorithm, especially there exist lots of channel paths.
	
	Fig.~\ref{S_capability} shows the SCR versus the edge computing capability, i.e., $f_\textrm{E}$. The benefits of exploiting MAs over FPAs are further conformed, especially when the edge computing capability is sufficient. The SCRs of all schemes increase with the edge computing capability firstly and become stable afterwards. This is due to the fact that the improvement in edge computing capability permits more tasks to be handled by task offloading, yet restricted by the WPT efficiency and channel state for task offloading. Therefore, for cost saving, it is crucial to take into account the network topology and equip the edge server with proper computing capability. Moreover, the partial schemes outperforms the counterpart offloading-only schemes because the partial schemes generally permit local computing and task offloading in parallel.
	
	Fig.~\ref{S_complexity} depicts the SCR versus the complexity of the tasks, $\varphi_k$. The SCRs of all schemes decrease as $\varphi_k$ increases, because the energy required to process 1-bit raw task via local computing is amplified and the time consumption via the edge computing becomes longer. The partial scheme outperforms the counterpart offloading-only schemes, which further validates the benefits of optimizing offloading strategies at the WDs. 

	\section{Conclusion}
	\label{section_conclusion}	
	In this paper, we have investigated three types of MA positioning configurations for the MA-enhanced WP-MEC systems, aiming to balance the enhancement of performance against the augmentation in implementation intricacy cased by the MA positioning optimization. For practical purposes, A non-linear EH model has been applied to characterize the power conversion of EH circuit at each WD. In addition, due to the practical hardware limitations, the finite edge computing capability has been considered. The SCR maximization problems for the three cases with different MA positioning configurations have been formulated by jointly optimizing the time allocation, positions of MAs, energy beamforming matrix, receive combing vectors, and offloading strategies of WDs. The dynamic MA positioning optimization has been studied at first, and the efficient AO framework with the PSO-VLS algorithm has been proposed. Then, the PSO-VLS algorithm has been extended to optimize the semi-dynamic and static MA positioning. Numerical results have revealed the superiority of exploiting MAs in WP-MEC systems over the conventional FPAs. Furthermore, analytical results have demonstrated that compared to the standard PSO algorithm, the PSO-VLS algorithm can jump from local optima.
	
	Finally, we discuss some future research directions. Significantly, it is optimistic that the positions of MAs can be adjusted freely within a given region. In practice, the motion control of the stepper motor is discrete with finite precision. Thus, the given region is quantized in several paces, leading to non-convex mixed integer programming (MIP). To tackle this intractable issue, the novel deep reinforcement learning (DRL)-based algorithms can be applied to optimize the resource allocation and offloading strategy in the MA-enhanced WP-MCE systems. Moreover, in MA-enhanced systems, it is impractical to acquire the instantaneous CSI due to the latency caused by the MA array movement. In this case, only statistical CSI is available, which may bring a great challenge to the MA positioning optimization.

	\bibliography{mybibliography}

\end{document}